\documentstyle[epsfig]{article}

 \makeatletter
 \def\section{\@startsection {section}{1}{\z@}{3.5ex plus 1ex minus
    .2ex}{2.3ex plus .2ex}{\sc }}
 \def\subsection{\@startsection{subsection}{2}{\z@}{3.25ex plus 1ex minus
   .2ex}{1.5ex plus .2ex}{\small \sc }}
 \def\subsubsection{\@startsection{subsubsection}{2}{\z@}{3.25ex plus 1ex minus
   .2ex}{1.5ex plus .2ex}{\small \sc }}
  \def\appendix{\par\clearpage
  \setcounter{section}{0}
  \setcounter{subsection}{0}
  \@addtoreset{equation}{section}
  \def\@sectname{Appendix~}
  \def\theequation{\thesection.\arabic{equation}}
  \def\thesection{\Alph{section}}}
\makeatother

\newcommand{\be}{\begin{equation}}
\newcommand{\ee}{\end{equation}}
\newcommand{\bea}{\begin{eqnarray}}
\newcommand{\eea}{\end{eqnarray}}
\newcommand{\noi}{\noindent}
\newcommand{\pss}{\protect\scriptscriptstyle}
\newcommand{\pst}{\protect\textstyle\scriptscriptstyle}
\newcommand{\f}{\frac}
\newcommand{\dfp}{\Delta^{\rm\pss FP}}
\newcommand{\dfpd}{\Delta^{{\rm\pss FP}\,\dagger}}
\newcommand{\pfp}{D^{\rm \pst FP}}
\newcommand{\rfp}{R^{\rm \pst FP}}
\newcommand{\umin}{U^{\rm\pst min}}
\newcommand{\xb}{x_{\pst b}}
\newcommand{\yb}{y_{\pst b}}
\newcommand{\ap}{\alpha^{\prime}}

\begin{document}
\thispagestyle{empty}
\parskip=12pt
\raggedbottom

\def\mytoday#1{{ } \ifcase\month \or
 January\or February\or March\or April\or May\or June\or
 July\or August\or September\or October\or November\or December\fi
 \space \number\year}
\noindent
\hspace*{7cm} BUTP-97/36\\
\hspace*{7cm} UNIGRAZ-UTP-02-02-98\\
\vspace*{1cm}
\begin{center}
{\LARGE Properties of the Fixed Point Lattice Dirac Operator in the
Schwinger Model.}\footnote{Work supported by Fonds zur F\"orderung der
Wissenschaftlichen Forschung in \"Osterreich - Austria, Project P11502-PHY,
and Ministerio de Educaci\'on y Cultura - Spain, grant PF-95-73193582.}

\vspace{1cm}

{\bf Federico Farchioni}\\
\vspace{0.2 cm}
Institute for Theoretical Physics \\
University of Graz \\
Universit\"atsplatz 5, A-8010 Graz, Austria\\
\vspace{0.5 cm}
and\\
\vspace{0.5 cm}
{\bf Victor Laliena}\\
\vspace{0.2 cm}
Institute for Theoretical Physics \\
University of Bern \\
Sidlerstrasse 5, CH-3012 Bern, Switzerland

\vspace{0.5cm}

\mytoday \\ \vspace*{0.5cm}

\nopagebreak[4]

\begin{abstract}
We present a numerical study of the properties of the Fixed Point
lattice Dirac operator in the Schwinger model. We verify the theoretical
bounds on the spectrum, the existence of exact zero modes with definite
chirality, and the Index Theorem. We show by explicit computation that it is
possible to find an accurate approximation to the Fixed Point Dirac operator
containing only very local couplings.
\end{abstract}

\end{center}
\eject

\section{\bf Introduction}

Lattice QCD is a tool to perform (in principle exact) non-perturbative
computations of physical quantities in problems where the strong
interaction effects are important, by means of numerical simulations.
Yet, available computer
resources do not allow simulations on lattices extended over a large enough
physical size (to keep the finite size effects under control) and with small
enough lattice spacing (to make the non-physical cut off dependence
negligible).

In the pure gauge sector, together with the contamination of the physical
results by lattice artifacts, we face up to an unpleasant situation: the
difficulties in studying the topological effects, a genuinely
non-perturbative problem.
At the classical level, it is possible to define a lattice topological
charge operator which, as in the continuum, assumes integer values
on all lattice configurations, the L\"uscher's~\cite{lue}
``geometrical'' definition. Unfortunately, at the quantum level
this operator displays a singular continuum limit, since it assigns
non-zero topological charge to lattice configurations extended over
a size which scales with the lattice spacing
(these configurations are called in the literature ``dislocations'').
An alternative approach~\cite{ftd} consists in defining the lattice topological
charge operator in terms of a local density, obtained from the
naive discretization of the continuum operator. At the classical level,
the resulting lattice operator, though not integer-valued,
has the right continuum limit.
At the quantum level, its matrix elements are connected to those of the
continuum by non trivial renormalizations~\cite{cdp}.
This approach, relying only on first principles of Quantum Field Theory,
is theoretically sound, but it has the inconvenience of the evaluation
of the renormalizations, a theoretical uncertainty being introduced
as a consequence in the lattice determinations.

Among the cut off effects, those in the fermion sector
are specially bothersome.
The Nielsen-Ninomiya theorem~\cite{nn} states the
impossibility of finding a local
lattice action which is chiral invariant and describes only one species fermion
in the continuum limit. A related problem is the impossibility of reproducing
the right chiral anomaly in the continuum limit with chirally symmetric
lattice actions. Since locality is of utmost importance for practical
purposes, any lattice action for QCD must necessarily
violate chiral symmetry.
This violation is so strong for the simplest discretizations of the fermion
action (``Wilson action'', ``clover action'') that all the chiral
properties of the continuum Dirac operator are lost on the lattice, although
they should be recovered in the continuum limit.
An example of these properties is, at the classical level, the existence
of zero modes with definite helicity,
and their relation to the topological charge of the background gauge
configuration through the Atiyah-Singer formula (Index Theorem)~\cite{as}:

\be
Q\;=\;n_{\pst L}\:-\:n_{\pst R}\;\;,
\label{it}
\ee

\noi
where $Q$ is the topological charge and $n_{\pst L}$ and $n_{\pst R}$ are
respectively the
number of the left- and right-handed zero modes of the Dirac operator.

At the quantum level, other important diseases are induced by the
violation of chiral the symmetry; an example is the necessity of a fine tuning
of the parameters to recover the (spontaneously broken) chiral symmetry
in the continuum limit (namely to get massless pions,
chiral Ward identities, etc.).

The freedom of choosing the lattice action among all possibilities
in the universality class of the proper renormalized theory can be exploited
in order to reduce the cut off effects at the price of rising the number
of couplings involved in the action and their extension in space-time.
The reward consists in more accurate physical results from
simulations on coarse lattices. The drawback is of course the slowing
down of the simulation procedure, due to the increased complexity
of the action. Whether all this can be of practical or only academic relevance
depends on the effective improvement of the accuracy of the results
for a fixed cost in computational time.

It follows as a consequence of the Wilson Renormalization Group (RG) ideas
that there exist actions which give exact physical answers no matter how
coarse the lattice is, the perfect actions~\cite{hlec}.
They are located on the renormalized trajectory of
any RG transformation. The first step toward perfect actions is represented by
the so called fixed point (FP) actions, which are perfect at the classical
level: this means, for asymptotically free theories, in the limit
$g\rightarrow 0$, where $g$ is the (asymptotically free) coupling constant.
It was shown by Hasenfratz and Niedermayer~\cite{hn0} that the
determination of the FP action of an asymptotically free theory
is a feasible objective, since it requires only the solution of a
classical minimization problem.
The FP action is by construction perfect at the classical level,
and as a consequence cut off effects are absent in the tree-level
term of the perturbative expansion of the spectral quantities of the theory;
for some specific observables, one-loop cut off effects
are absent~\cite{fphn,fv}, or strongly suppressed
with respect to a standard discretization~\cite{hn1l}.

In the context of the SU(3) Yang-Mills theory, the properties of
the FP action are by now well known from a theoretical
point of view~\cite{hnymp}, and they have been extensively tested in Monte
Carlo simulations~\cite{hnymn,aless,aless2}, with the result of
very small cut off effects for spectral quantities
even at moderate correlation lengths.
Also, the ideas of
the RG allow to build a well-defined
topological charge operator on the lattice, the FP topological charge operator,
which assumes integer values on all gauge configurations and is not affected
by topological defects~\cite{bbhn,dfp}.

The next step, the inclusion of quarks, is yet in a preliminary
phase~\cite{dhhkn,fermion}.
It is theoretically clear~\cite{has,hln} that many of the diseases of the
simplest fermion actions produced by the violation of the chiral symmetry are
cured by working with a FP action, whose (obligatory) chiral asymmetry is so
mild that it has no effect in the physical results: the classical
statements like the existence of zero modes of topological origin and the Index
Theorem hold~\cite{hln}; at the quantum level, chiral symmetry is recovered at
zero bare quark mass, without any tuning of an additional parameter,
the currents get no renormalization, and there is no mixing between
composite fermion operators corresponding to different chiral
representations~\cite{haslast}.
This properties are deduced using the Ginsparg-Wilson (GW) relation~\cite{gw},
which expresses  the mildest breaking of the chiral symmetry for a ``legal''
fermion action; this relation is satisfied by any FP Dirac operator.
Very recently, L\"uscher has pointed out that fermion actions verifying the
GW relation have an exact continuous symmetry, which can be interpreted as
a lattice implementation of the chiral symmetry without any contradiction with
the Nielsen-Ninomiya theorem~\cite{luescher}.

It is worth to observe in this context that a solution of
GW equation was obtained independently from the RG techniques.
This solution is a product of the overlap formalism and shares with the
FP Dirac operator its gentle chiral properties~\cite{neuberger}.

In this paper we want to verify the classical properties of the FP action
in the fermion sector in a model simpler than QCD: the Schwinger model.
The work is analytical for the pure gauge sector and numerical
for the fermion sector.
We investigate in particular how the approximation of the FP action
with a finite-range fermion matrix affects its (classically perfect)
properties.
The paper is organized as follows. In Sec. 2 we recall the
recursion relation obeyed by the FP Dirac operator,
enlightening some properties of its classical solutions
and bounds on the spectrum; in particular, we formulate
the lattice version of the Index Theorem.
In Sec. 3 we describe  some
topological properties of the continuum Schwinger model relevant for the
subsequent work and define the RG transformation which generates the
FP action under study, discussing the analytical results of
the pure-gauge sector. Sec. 4 is devoted to the discussion of the
numerical results concerning the spectrum of the FP Dirac operator.
The paper ends with a summary of the conclusions in Sec. 5.

\section{\bf The Fixed Point lattice Dirac operator}
\label{sec:fpdo}

\subsection{Recursion relations}

Let us recall schematically how the FP action for asymptotically free gauge
theories can be computed; for further details, we refer the reader to the
literature~\cite{hnymp, hlec}. Consider a gauge theory, whose lattice action
can be written as

\be
S[\,U,\psi,\bar\psi\,]\;=\;\beta\,S_{\rm\pst G}(U)\:+\:
\sum_{n,m}\,\bar\psi_n\,\Delta_{n m}(U)\,\psi_m \, ,
\ee

\noi
where $S_{\rm\pst G}(U)$ is a lattice action for the pure gauge part, for
example the
Wilson single plaquette action, and $\Delta(U)$ is a proper discretization of
the continuum Dirac operator. On the lattice, the latter operator has
the form of a matrix with space-time, Dirac and color indices, which
in the above formula have been indicated in a collective notation.
Since we do not want replica fermions in the
spectrum, $\Delta(U)$ cannot be chiral symmetric, i.e. it must mix-up
left- and right-handed spinor components.

We define a physically equivalent action\footnote{With the same long distance
physics.} on a lattice with doubled lattice spacing through a RG
transformation:

\bea
&&\hspace*{-1 cm}
\exp\,\left[\,-\,\beta^\prime\,S_{\rm\pst G}^\prime(U^\prime)\:-\:
S_{\rm\pst F}^\prime(\bar\psi^\prime,\psi^\prime,U^\prime)\,\right]\;=
\nonumber \\
& &\hspace*{1 cm} \int\,\left[dU\,d\bar\psi\,d\psi\right]\,\exp\,\left[
-\,\beta\,S_{\rm\pst G}(U)\:-\:\bar\psi\,\Delta(U)\,\psi\:-\:p\,{\cal
K}_{\rm\pst G}(U^\prime,U)
\right. \nonumber \\
& &\hspace*{1 cm}
\left. -\:\kappa_{\rm \pst F}\,{\cal K}_{\rm\pst
F}(\bar\psi^\prime,\psi^\prime,\bar\psi,\psi,U)
\,\right]\, , \label{rgtdef}
\eea

\noi
where the primed fields are the degrees of freedom on the coarser lattice,
defined through the gauge invariant kernels \ ${\cal K}_{\rm\pst G}$ \ and \
${\cal K}_{\rm\pst F}$.
The parameters $p$ and $\kappa_{\rm \pst F}$ can be chosen
arbitrarily\footnote{The
choice is somehow restricted  by the request that the RG transformation
converges to a FP when iterated infinitely many times.}; in particular,
we can take $p=\beta\,\kappa_{\rm\pst G}$, where $\kappa_{\rm\pst G}$ is a
fixed parameter.
Note that in general the fermion action is
not quadratic in the fermion fields after a RG transformation.

In the class of theories under interest (including asymptotically
free non-abelian gauge theories and the Schwinger model),
the critical surface (where the continuum is attained) is at $\beta=\infty$.
The iteration of the RG transformation starting on this surface
converges to a FP.
When $\beta\rightarrow\infty$ the integral on the gauge
degrees of freedom in the r.h.s. of Eq.~(\ref{rgtdef}) is saturated by the
saddle point configuration; the solution of the recursion is then:

\bea
& & S_{\rm\pst G}^\prime(U^\prime)\;=\;\min_{\{ U \} }\,\left[S_{\rm\pst
G}(U)\:+\:
\kappa_{\rm\pst G}\,{\cal K}_{\rm\pst G} (U^\prime,U)\,\right] \, ,
 \label{sapo} \\
& & S_{\rm\pst F}^\prime (\bar\psi^\prime,\psi^\prime,U^\prime)\;=\;-\ln\,
\int\,\left[d\bar\psi\,d\psi\right]\,\exp\,\left[\,
-\:\bar\psi\,\Delta(U^{\rm\pst min})\,\psi \right. \nonumber \\
& &\left.\;\;\;\;\;\;\;\;\;\;\;\;\;\;\;\;\;\;\;\;\;\;\;\;\;\;\;\;\;\;\;\;\;\;\;\;
-\:\kappa_{\rm \pst F}\,{\cal K}_{\rm\pst F}(\bar\psi^\prime,\psi^\prime,
\bar\psi,\psi,U(U^{\rm\pst min}))\,\right]\, , \label{sapof}
\eea

\noi
where $\umin$ is the fine gauge field configuration which minimizes
the r.h.s. of Eq.~(\ref{sapo}). It depends on the coarse
configuration $U^\prime$. We have also $\beta^\prime=\beta=\infty$.
In this limit the problem is equivalent to a classical minimization problem
plus a Grassmann integration.

If the fermion kernel is quadratic in the fermion fields,
the fermion action at $\beta=\infty$ remains quadratic after a RG
transformation, as is evident from Eq.~(\ref{sapof}). We write in general

\be
{\cal K}_{\rm\pst F} \;=\;
\sum_{\xb}\,\left(\,\bar\psi^\prime_{\xb}\:-\:\sum_x\,\bar\psi_x\,
\omega_{x \xb}^\dagger\,\right)\,\left(\,\psi^\prime_{\xb}\:-\:
\sum_y\,\omega_{\xb y}\,\psi_y\,\right)\;\;,
\label{rra}
\ee

\noi
where $\xb$ and $x, y$ label the sites in the coarse and the fine lattices
respectively; $\omega$ is a matrix in space-time and color indices,
depending on the gauge field $U$ in such a way
that gauge invariance is preserved, and normalized according to:
$\sum_x\omega_{\xb x}=2^{\f{d-1}{2}}$. The adjoint operation denoted
with the symbol $\dagger$ acts on all indices,
including space-time.

The FP action is defined as

\be
S^{\rm \pss FP}\;=\;\beta\,S^{\rm \pss FP}_{\rm\pst
G}(U)\:+\:\bar\psi\,\Delta^{\rm \pst
FP}(U)\,\psi\, ,
\label{rrp}
\ee

\noi
where $S^{\rm \pss FP}_{\rm\pst G}$ and $\Delta^{\rm \pss FP}$ are the
self-reproducing
solutions of Eq.~(\ref{sapo}) and  Eq.~(\ref{sapof}) respectively.
The recursion relation for the FP action of the pure gauge sector
and for the FP Dirac operator read respectively:

\be
S_{\rm\pst G}^{\rm \pss FP}(U)\;=\;S_{\rm\pst G}^{\rm \pss FP}(\umin (U))\:+\:
\kappa_{\rm\pst G}\,{\cal K}_{\rm\pst G} (U,\umin (U))\;\; , \label{gfp}
\ee
\be
\dfp_{\xb \yb}(U)\;=\;\kappa_{\rm \pst F}\,\delta_{\xb
\yb}\:-\:\kappa_{\rm \pst F}^2\,
\sum_{x y}\,\omega_{\xb x}\,\left(\f{1}{\dfp\,+\,\kappa_{\rm \pst F}\,
\omega^\dagger\,\omega}\right)_{x y}\,
\omega^\dagger_ {y \yb}\, , \label{ffp}
\ee

\noi
where $\umin (U)$ indicates
the lattice gauge configuration defined on the fine lattice realizing the
minimum in the r.h.s of Eq.~(\ref{gfp}) for a fixed coarse lattice
configuration $U$. Notice that, although not explicitly displayed, all the
matrices in the r.h.s. of Eq.~(\ref{ffp}) depend on the coarse
gauge configuration $U$ through $\umin (U)$.

The recursion relation for the inverse Dirac operator $D\equiv\Delta^{-1}$
- the fermion propagator in the background of the gauge configuration $U$ -
is linear and therefore simpler and often more convenient\footnote{It
should be noted however that, due to possible zero modes of the Dirac operator,
its inverse does not always exist.}:

\be
\pfp_{\xb \yb}(U)\;=\;\sum_{x y}\,\omega_{\xb x}(\umin)\,\pfp_{x y}(\umin)
\,\omega^\dagger_{y \yb}(\umin)\:+\:\f{1}{\kappa_{\rm \pst F}}\,\delta_{\xb
\yb}\;\;.
\label{prorec}
\ee

\noi

\subsection{Fixed point operators}
\label{subs:fpo}

Suppose $O(U)$ to be some lattice discretization of a continuum operator
depending only on the gauge fields.
The RG transformation for such operator reads~\cite{peda}:

\bea
&&\hspace*{-1 cm}
\lambda\,O^\prime(U^\prime)\,\exp\,\left[\,-\,\beta^\prime\,S_{\rm\pst
G}^\prime(U^\prime)\:-\:
S_{\rm\pst F}^\prime(\bar\psi^\prime,\psi^\prime,U^\prime)\,\right]\;=
\nonumber \\
& &\hspace*{1 cm} \int\,\left[dU\,d\bar\psi\,d\psi\right]\,O(U)\,\exp\,\left[
-\,\beta\,S_{\rm\pst G}(U)\:-\:\bar\psi\,\Delta(U)\,\psi\:-\:p\,{\cal
K}_{\rm\pst G}(U^\prime,U)
\right. \nonumber \\
& &\hspace*{1 cm}
\left. -\:\kappa_{\rm \pst F}\,{\cal K}_{\rm\pst
F}(\bar\psi^\prime,\psi^\prime,\bar\psi,\psi,U)
\,\right]\;\; , \label{oprgtdef}
\eea

\noi
where the eigenvalue $\lambda$ is determined by dimensional considerations.
This relation preserves (apart from a trivial multiplicative factor)
the matrix elements of the operator $O(U)$
if, of course, also the action is changed accordingly.

In the $\beta\rightarrow\infty$ limit, if the action is at the FP,
the above equation reduces to:

\be
\lambda\,O^\prime(U^\prime)\;=\;O(\umin (U^\prime))\;\;. \label{oprec}
\ee

\noi
Eq.~(\ref{oprec}) has a FP in the operator $O^{\rm \pss FP}(U)$ formally
defined by the following  limit:

\be
O^{\rm \pss FP}(U)=\lim_{n\rightarrow\infty}(1/\lambda)^n O(U^{\pst(n)}
(U))\;\;,
\label{opfp}
\ee

\noi
where $U^{\pst(n)} (U)$ is obtained by applying $n$ times
- in a recursive way - Eq.~(\ref{gfp}) starting from the coarse configuration
$U$:

\be
U^{\pst (i)}\;=\;\umin(U^{\pst (i-1)})\;\;\;\;\; {\rm and} \;\;\;\;\;
U^{\pst (0)}\;\equiv\;U \, . \label{confrec}
\ee

As is evident from the definition~(\ref{opfp}), the FP operator
$O^{\rm \pss FP}$ is the solution of the FP equation:

\be
\lambda\,O^{\rm \pss FP}(U)\;=\;O^{\rm \pss FP}(\umin(U))\;\;.
\ee

$O^{\rm \pss FP}(U)$ represents the classical perfect lattice operator, in the
sense that its classical properties are the same of the corresponding
continuum operator.

A case of particular interest is that of the topological charge operator.
Indeed, the FP charge operator $Q^{\rm \pss FP}(U)$ allows a theoretically
sound definition of the topology on the lattice, since it assumes integer
values and it is not affected by short-range fluctuations
(``dislocations'')~\cite{bbhn,dfp}.

\subsection{Classical solutions and the Index Theorem}

In analogy with the pure-gauge sector, where there exist
exact classical solutions of the lattice equations of motion having the main
properties of the continuum solutions (e.g. scale invariance)~\cite{hnymp},
it is also possible to prove~\cite{has} the existence of exact solutions
of the lattice FP Dirac equation, given  by the zero modes
of the FP Dirac operator.
The arguments given here follow~\cite{has}; see also~\cite{hln}
for an alternative approach.

Suppose $f^{\pst (c)}_{\xb}$ to be a solution of the FP Dirac equation
in the background of some coarse configuration $U^{\pst (c)}$
(Dirac and color indices are understood):

\be
\sum_{\yb}\,\dfp_{\xb \yb}(U^{\pst (c)})\, f^{\pst (c)}_{\yb} \;=\;0\;\;.
\label{code}
\ee

\noi
Using (\ref{ffp}) it is easy to verify that

\be
f^{\pst (f)}_x\;=\;\sum_{y y_B}\,\left(\f{1}{\dfp(\umin)
\,+\,\kappa_{\rm \pst F}\,\omega^\dagger(\umin)\omega(\umin)}\right)_{x y}\,
\omega^\dagger_{y y_B}(\umin)\,f^{\pst (c)}_{y_B}
\label{fisol}
\ee

\noi
is also a solution of the Dirac equation in the fine lattice with the
background configuration $\umin(U^{\pst (c)})$:

\be
\sum_y\,\dfp_{x y}(\umin(U^{\pst (c)}))\,f^{\pst (f)}_y\;=\;0
\label{fide}
\ee

Conversely, let $f^{\pst (f)}_x$ be a solution of the Dirac equation in the
fine lattice, with the background configuration $\umin(U^{\pst (c)})$. Then

\be
f^{\pst (c)}_{\xb}\;=\;\sum_x\,\omega_{\xb x}(\umin)\,f^{\pst (f)}_x
\label{cosol}
\ee

\noi
is a solution of the Dirac equation in the coarse lattice with the background
configuration $U^{\pst (c)}$.

Eqs.~(\ref{fisol}) and (\ref{cosol}) establish a one-to-one correspondence
between the zero modes of the FP Dirac operator in
the background of the coarse gauge configuration $U^{\pst (c)}$
with those of the same operator, but considered now
in the background of the fine gauge configuration $\umin (U^{\pst (c)})$.
From Eq.~(\ref{cosol}) we see that two related zero modes must have
the same chirality, since $\omega$ does not depend on Dirac indices.
By iteration, one can relate in the same way  the zero modes
of $\dfp_{x y}(U^{\pst (c)})$ to those of
$\dfp_{x y}(U^{\pst (n)} (U^{\pst (c)}))$, where
$U^{\pst (n)} (U^{\pst (c)})$ is defined as in Eq.~(\ref{confrec}).
In the limit of infinite iterations $n\rightarrow\infty$,
$U^{\pst (n)} (U^{\pst (c)})$ converges to a continuous configuration.
The topological charge of such configuration is given by

\be
Q=\lim_{n\rightarrow\infty} Q^L(U^{\pst (n)} (U^{\pst (c)}))\;\;,
\ee

\noi
where $Q^L$ indicates any proper lattice discretization
of the topological charge
operator. The r.h.s. of the above equation is just the definition
of the FP topological charge
of the coarse lattice configuration $U^{\pst (c)}$,
$Q^{\rm \pss FP}(U^{\pst (c)})$ (see Eq.~(\ref{opfp}); in the case of
the topological charge, a dimensionless operator, $\lambda=1$).
For the limiting continuous configuration the Index Theorem holds, since
$\dfp_{x y}(U^{\pst (n)})$ converges to the continuum operator
when $n\rightarrow\infty$. Using the above established
one-to-one correspondence between
the zero modes, and the fact that this correspondence preserves chirality,
we arrive to the conclusion that~\cite{has,hln}:
\begin{itemize}
\item[i)] the FP Dirac operator $\dfp(U)$ possesses {\em exact} zero modes
of topological origin;
\item[ii)] these zero modes have definite chirality and satisfy the
Index Theorem, if the lattice topological charge is the FP topological
charge $Q^{\rm \pss FP}(U)$ of the background lattice gauge
configuration $U$.
\end{itemize}

\subsection{The Ginsparg-Wilson relation}

Let us show now how, forced by the Nielsen-Ninomiya theorem, the FP Dirac
operator violates the chiral symmetry~\cite{has,hln}.
Consider the n-th iteration of Eq.~(\ref{prorec}):

\be
\pfp_{\xb \yb}(U)\;=\;\sum_{x y}\,\Omega^{\pst (n)}_{\xb x}(U)\,
\pfp_{x y}(U^{\pst (n)})\,\Omega^{{\pst (n)}\,\dagger}_{y \yb}(U)\:+\:
R^{\pst (n)}_{\xb \yb}(U)\, ,
\label{rrpiter}
\ee

\noi where

\be
R^{\pst (n)}_{\xb \yb}(U)\;=\;\frac{1}{\kappa_{\rm \pst F}}
\left(\delta_{\xb \yb}\:+\:\sum_{i=1}^{n-1}\,
\sum_x\,\Omega^{\pst (i)}_{\xb x}\,\Omega^{{\pst (i)}\,\dagger}_{x \yb}(U)
\right)
\label{rdef}
\ee

\noi and

\be
\Omega^{\pst (i)}_{\xb x}(U)\;=\;\sum_{x_1\ldots x_{i-1}}\,
\omega_{\xb x_1}(U^{\pst (1)})\,\omega_{x_1 x_2}(U^{\pst (2)})\ldots
\omega_{x_{i-1} x}(U^{\pst (i)})\, ,
\ee

\noi
with $U^{\pst (i)}(U)$ defined as in Eq.~(\ref{confrec}).

The background configuration $U^{\pst (n)}$ in Eq.~(\ref{rrpiter})
becomes arbitrarily smooth when increasing $n$. Therefore,
when $n\rightarrow \infty$ the propagator in the r.h.s. of (\ref{rrpiter})
converges to the continuum propagator, which is chiral invariant. Since
$\Omega^{\pst (n)}$ commutes with $\gamma_5$ (we recall that $\omega$
does not act on Dirac indices), we have~\cite{has}

\be
\{\,\pfp_{\xb \yb},\, \gamma_5\,\}\;=\;2\,\gamma_5\,
R^{\rm \pst FP}_{\xb \yb}(U)\, ,
\label{gwp}
\ee

\noi
where $R^{\rm \pst FP}$ the limit $n\rightarrow\infty$ of $R^{\pst (n)}$.
Equivalently, we have

\be
\{\,\dfp_{\xb \yb},\,\gamma_5\,\}\;=\;2\,\left(\,
\dfp\,\gamma_5\,R^{\rm \pst FP}\,\dfp\,\right)_{\xb \yb}\;\;.
\label{gwa}
\ee

\noi
The hermitian matrix $\rfp_{\xb \yb}$ is local and its spectrum is
bounded~\cite{hln}:

\be
1\;\leq\;\kappa_{\rm \pst F}\,\f{(v,\rfp
v)}{(v,v)}\;\leq\;\f{1}{1-q}\;\;,\;\;\;\;0\leq q<1\;\;;
\label{rbound}
\ee

\noi
the lower bound comes trivially from Eq.~(\ref{rdef}), while the upper bound
(specifically the constant $q$) is RG-transformation-dependent.
In the case of the overlapping-symmetric fermion RG
transformation~\cite{dhhkn,kun}, and its gauge-invariant extension
for the interacting theory~\cite{fv} here considered (see the following),
one has $q=1/2$.

Ginsparg and Wilson proved~\cite{gw} that
any lattice discretization of the Dirac
operator which verifies Eqs. of the form~(\ref{gwp})-(\ref{gwa}) -
it is crucial the locality of the matrix $\rfp_{\xb \yb}$ -
reproduces the correct axial anomaly in the continuum limit. We refer
Eqs.~(\ref{gwp})-(\ref{gwa}) as the Ginsparg-Wilson relation (GWR).
It expresses the explicit violation of chiral
symmetry by the lattice regulator. The fact that this violation
takes the form of the GWR ensures that many important chiral
properties of the continuum Dirac operator hold also for its FP discretization.

For instance, using the GWR the Index Theorem has been proven in
Ref.~\cite{hln} directly on the lattice, without any reference to the
Theorem of the continuum, finding also the following relation
between the FP topological charge and the FP Dirac operator:

\be
Q^{\rm \pst FP}(U)\;=\;-\f{1}{2}\,{\rm
tr}\left(\gamma_5\,\{\dfp(U),\,\rfp(U)\}
\right)\, .
\label{top}
\ee

\subsection{About the spectrum}

The GWR strongly constrains the spectrum of the FP Dirac operator.
Consider first the zero modes. Let us denote by $v$ a vector column
which contains the spatial, Dirac and color indices.

\noi {\em Property:} \\
$v$ is a zero mode of $\dfp$ if and only if $\gamma_5\,v$ is. Therefore, the
zero modes can be chosen eigenstates of $\gamma_5$.

The proof is trivial using (\ref{gwa}):

\be
\dfp\,\gamma_5\,v\;=\;-\gamma_5\,\dfp\,v\:+\:\{\dfp,\,\gamma_5\}\,v\;=\;0\;\;.
\ee

In what follows we shall prove that the FP Dirac operator is bounded, and
we shall find the explicit bound on the spectrum. First, we notice that the
FP Dirac operator verifies the hermiticity condition

\be
\dfpd\;=\;\gamma_5\,\dfp\,\gamma_5\, ,
\label{hc}
\ee

\noi
since it is preserved by the recursion relation (\ref{rra}). The hermiticity
property, which also holds for the Wilson Dirac operator, implies that the
eigenvalues of $\dfp$ are either real or comes by pairs of complex conjugate
numbers. It also implies that if $v$ is an eigenvector corresponding to a non
real eigenvalue $\lambda$ ($\lambda\neq\lambda^*$) then

\be
(v,\gamma_5 v)\;=\;0\;\;.
\ee

Multiplying on the left by $\gamma_5$ both sides of (\ref{gwa}) and using
(\ref{hc}), we obtain:

\be
\dfp\: +\: \dfpd\;=\;2\,\dfpd\,\rfp\,\dfp\, .
\label{sar}
\ee

\noi
Let $v$ be an eigenvector of $\dfp$ with eigenvalue $\lambda$, normalized to
unity: $(v,v)=1$. Multiplying both sides of (\ref{sar}) by $v^\dagger$ on the
left and by $v$ on the right, we find:

\be
\lambda\:+\:\lambda^*\;=\;2\,(v,\rfp v)\,
|\lambda|^2\;\; .
\ee

\noi
The above equation, in conjunction with the bounds~(\ref{rbound})
for the $\rfp$ matrix, allows to state analogous bounds for
the spectrum of the FP Dirac operator~\cite{hln}:

\be
\f{2}{\kappa_{\rm \pst F}}\,|\lambda|^2\;\leq\;\lambda\:+\:\lambda^*\;
\leq\;\f{2}{(1-q)\,\kappa_{\rm \pst F}}\:|\lambda|^2\;\; .
\label{bound}
\ee

\noi
These inequalities mean that the spectrum of the FP Dirac operator lies in the
complex plane inside a circle of radius $\kappa_{\rm \pst F}/2$ centered at
$(\kappa_{\rm \pst F}/2, 0)$, and outside a circle of
radius $(1-q)\,\kappa_{\rm \pst F}/2$ centered at $((1-q)\,\kappa_{\rm \pst
F}/2, 0)$.

\section{\bf The case of the Schwinger model}
\label{sec:schw}

\subsection{Topological properties of Schwinger model}

The Schwinger model is a good laboratory to test all these ideas, since it is
much simpler than four-dimensional non-abelian gauge theories, but
still its gauge sector has non-trivial topology.
The topological charge of a gauge configuration $A_\mu$ is given by:

\be
Q\;=\;\f{e}{2\pi}\,\int d^2x\, F_{\pst 12}\;\;,
\label{ccha}
\ee

\noi
where: $F_{\pst 12}=\partial_{\pst 1} A_{\pst 2} -
\partial_{\pst 2} A_{\pst 1}$ and $e$ is the electric charge.
For the theory defined on a torus $T^2$ ($\,x_\mu\in[0,L_\mu],\,\mu=1,2\,$),
topologically non-trivial
configurations must necessarily have a jump at the boundary.
Of course, gauge invariant functionals of the gauge fields must be
continuous.
The simplest topologically non-trivial gauge configurations
are solutions of the classical equations of motion, therefore with constant
$F_{\pst 12}$; in this case a possible expression for the gauge fields
is~\cite{vanbaal}:

\be
A^{{\rm \pss inst}}_{\pst 1}(x)\;=\;-\,\Phi\,x_{\pst 2}\,
,\;\;\;\;\;\;\;\;\;\;\;\;\;
A^{\rm \pss inst}_{\pst 2}(x)\;=\;0\, ,
\;\;\;\;\;\;\;\;\;\;\;\;\;\;\; F_{\pst 12}\;=\;\Phi\; .
\label{cinst}
\ee

The fields $A^{\rm \pss inst}_{\pst 1}(x_{\pst 2}=L_{\pst 2})$ and
$A^{\rm \pss inst}_1(x_{\pst 2}=0)$ can
be connected by a transition function (ensuring continuity for gauge-invariant
quantities) if $\Phi=2\pi Q/(L_1L_2e)$, with
%
%
$Q$ some integer number, corresponding to the topological charge
of the gauge configuration on the torus according to definition~(\ref{ccha}).

In perfect analogy to the case of QCD, also in the Schwinger model it is
possible to relate the chiral properties of the zero-modes of the massless
Dirac operator to the topological properties of the background gauge
configuration $A_\mu$. This relation is given by the Atiyah-Singer Index
theorem,
Eq.~(\ref{it}).
A peculiarity of two space-time dimensions is that:

\bea
Q(A)\;>\;0 \;\Rightarrow\; n_{\pst R}\; =\; 0\;\;,\nonumber\\
Q(A)\;<\;0 \;\Rightarrow\; n_{\pst L}\; =\; 0\;\;,
\label{eq:van}
\eea

\noi
the so called ``Vanishing Theorem''~\cite{van}.
So, in the case $Q(A)\neq 0$, the zero modes have all the same chirality,
given by the sign of $Q(A)$.

\subsection{The pure gauge sector on the lattice}

To be able to study the topological properties of the Schwinger model on
the lattice, we discretize the theory in the compact formulation,
where the lattice gauge fields are described by angular variables
$\alpha_\mu(x)\in(-\pi,\,\pi]$ and, forced by the necessity
of preserving gauge invariance, the fermions are coupled to the gauge
fields through $U(1)$ elements of the form:

\be
U_\mu(x)\;=\;{\rm e}^{i\,\alpha_\mu(x)}\;\;.
\label{angular}
\ee

\noi
The lattice action must be a periodic function of $\alpha_\mu$.
Within this formulation, we can implement periodic boundary conditions
also for the topologically non-trivial gauge field configurations.
The lattice discretization of the continuum instanton
solution on the torus (\ref{cinst}) is~\cite{vs}:

\bea
\alpha^{\rm \pss inst}_{\pst 1}(x) &=&
- \Phi^{\pst L}\,x_{\pst 2}\;({\rm mod}\,2\pi)
\nonumber \\
\alpha^{\rm \pss inst}_{\pst 2}(x) &=& 0
\hspace{3.76 truecm} {\rm if}\;\;x_{\pst 2}=0,1,\ldots,N_{\pst 2}-2
\nonumber \\
\  &=& \Phi^{\pst L} N_{\pst 2}\,x_{\pst 1}\;({\rm mod}\,2\pi)
\hspace{1 truecm} \;\;{\rm if}\;\;x_{\pst 2}=N_{\pst 2}-1\, ,
\label{inst}
\eea

\noi
where $\Phi^{\pst L}=2\pi Q/(N_{\pst 1}N_{\pst 2})$ and
$[\theta\;({\rm mod}\,2\pi)]\,\in(-\pi,\,\pi]$.

To carry on our program of finding a FP lattice action we must define a RG
transformation. For the pure gauge part, we consider the following kernel
(the primed variables are those relative to the coarse lattice):

\be
{\cal K}_{\rm\pst G}(\alpha^\prime,\alpha)\;=\;
\sum_{x_b,\,\mu}\left([\alpha^{\prime}_\mu(x_b)\:-\:\Gamma_\mu(x_b)]\;
({\rm mod}\,2\pi)\,\right)^2\;\;,
\label{kern}
\ee

\noi
where $\Gamma_\mu(x_b)=[\,\alpha_\mu(x)\:+\:\alpha_\mu(x+\hat{\mu})\,]
({\rm mod}\;2\pi)\,$.


In the limit $\kappa_{\rm\pst G}\rightarrow\infty$, Eq.~(\ref{sapo}) reduces
to a problem of constrained minimum:

\be
S_{\rm\pst
G}^\prime(\alpha^{\prime})\;=\;4\,\min_{\{\alpha(x)\}}\,\left\{
S_{\rm\pst G}(\alpha)\:|\:\Gamma_\mu(x_b)\,=\,\alpha^{\prime}_\mu(x_b) \right\}
\;\; .  \label{oursapo} \\
\ee

\noi
The $4$ factor comes out since in two dimensions the electric charge is a
relevant parameter having the dimension of a mass and, therefore, the
coupling constant is trivially renormalized at the lowest (tree level)
order: $\beta^\prime=\beta/4$.

In the Appendix we proof that the FP of the above recursion
relation is given by:

\be
S_{\rm\pst G}^{\rm \pss FP}(\alpha)\;=\;\f{1}{2}\,\sum_x\,\left[\,
F^{\pst L}_{\pst 12}(x)\;({\rm mod}\;2\pi)\,\right]^2\;\;,
\label{man}
\ee

\noi
where $F^{\pst L}_{\mu\nu}(x)$ is the lattice field strength tensor:

\be
F^{\pst L}_{\mu\nu}\;=\;\alpha_\mu(x)\:+\:\alpha_\nu(x+\hat{\mu})\:-\:
\alpha_\mu(x+\hat{\nu})\:-\:\alpha_\nu(x)\;\;.
\ee

We have so obtained the Manton action~\cite{manton} of the U(1) theory.
This is not a surprise since it is known~\cite{manton2}
that the $U(1)$ pure gauge theory in
two dimensions is equivalent to the one dimensional quantum rotor, for which
the Manton action was proven~\cite{bbcw} to be classically perfect.
In \cite{arjan} it was pointed out that the string tension computed with the
Manton action shows perfect scaling.

In the Appendix we also report an explicit solution
(among all the possible gauge-equivalent ones) for
the fine configuration minimizing the r.h.s. of Eq.~(\ref{oursapo}),
$\alpha^{\rm \pst min}(\alpha^{\prime})$, giving trivially
through~(\ref{angular})
$\umin (U^{\prime})$.

The action in Eq.~(\ref{man}) can be rewritten in terms of
the link variables:

\be
S_{\rm\pst G}^{\rm \pss FP}(U)\;=\;\f{1}{2}\,\sum_x\,
\left(\,{\rm Im} \ln(U_{\pst 12}(x))\,\right)^2\;\;,
\label{uman}
\ee

\noi
where $U_{\pst 12}(x)$ represents the usual product of links around
the $1-2$ plaquette. The r.h.s. of (\ref{uman}) is not defined for the
so-called exceptional configurations\footnote{The term exceptional
configurations is used in a different context to denote those configurations
for which the inversion of the Dirac operator is numerically complicated by
the presence of nearly zero modes.},
which have $U_{\pst 12}(x)=-1$ for some $x$. These, however, have zero measure
in the path integral and can be ignored.
We see that the FP action, though ultra-local,
involving only first neighbor interactions, has a non-polynomial
dependence on the link variables. In particular, considering the expansion:

\be
\ln(U_{\pst 12})\;=\;-\sum_{k=1}^{\infty}\,\f{1}{k}\,(1-U_{\pst 12})^k\;\;,
\ee

\noi
we understand that the fixed point action takes contribution from monomials
of the link variables of arbitrary order, with a slow suppression
with the increasing order. This may suggest that, in the framework of
the non-abelian Yang-Mills theory, the choice to
parametrize the FP action by a finite set of operators
of loops of the link variables is definitely not the
most suitable, and other approaches should be dared~\cite{has}.

The lattice configuration of Eq.~(\ref{inst}) is a classical
solution for the FP action~(\ref{man})-(\ref{uman}). It is in general
expected that a FP action (which - we recall - is a classically perfect action)
reproduces the continuum action on classical solutions~\cite{hn0}.
This is indeed the case in the present context, since,
as it is clear from the expression~(\ref{man}):

\be
\beta\,S_{\rm\pst G}^{\rm \pss FP}(\alpha^{\rm \pss inst})\;=\;\f{2\pi^2\beta
Q^2}{N^2}\;\;;
\ee

\noi
exploiting the relation $\beta=1/e^2a^2$,
the r.h.s of the above equation can be rewritten in physical units as
$2\pi^2 Q^2/L^2e^2$, i.e. the continuum action of the
configuration~(\ref{cinst}).

To end the discussion of the pure gauge sector, we shall study the FP
topological charge. Using the expression given in the Appendix for
$\alpha^{\rm\pss min}(\alpha^{\prime})$, it is straightforward to prove
that the solution of the recursion relation:

\be
Q^{\rm \pss FP}(\alpha)\;=\;Q^{\rm \pss FP}(\alpha^{\rm\pss min}(\alpha))
\label{charec}
\ee

\noi
is given by (cfr.~\cite{bbcw}):

\be
Q^{\rm \pst FP}(\alpha)\;=\;\f{1}{2\pi}\sum_x\,F_{\pst 12}^{\pst L}(x)\;
({\rm mod}\,2\pi)\;\;.
\label{fpch}
\ee

\noi
The r.h.s. of the above equation is an integer, since
$\sum_x F_{\pst 12}^{\pst L}(x)=0$ due to the periodic
boundary conditions. Notice that the geometrical charge,

\be
Q^{\pss G}\;=\;\f{1}{2\pi}\,\sum_x\,{\rm Im\, ln}(U_{\pst 12})\; ,
\label{geocha}
\ee

\noi
coincides with the FP charge for non-exceptional
configurations\footnote{The FP charge fixes the prescription:
$\ln(-1)\!=\!i\pi$ in the case of an exceptional configuration.}.
Since exceptional configurations have zero measure in the path integral, we
see that the geometrical charge gives the correct continuum limit. This is
somehow different from what happens in the non-abelian theories, where the
geometrical charge is singular in the continuum limit due to the
dislocations~\cite{geo}.

\subsection{RG transformation for the fermions}

For the fermion sector, we choose the quadratic RG kernel (\ref{rra})
with the following matrix $\omega$~\cite{fv}:

\bea
\omega_{x_b x}(U)\:=\:\f{\sqrt{2}}{4}\left\{\delta_{x,\,2x_b}\:+\:\f{1}{2}
\sum_{\mu=\pm1,\pm2}
U_{\mu}(2x_b)\,\delta_{x,\,2x_b+\hat{\mu}}\right.\:+\;\;\;\;\;\;\;\;\;\;\nonumber\\
\left. \f{1}{4}\sum_{\mu=\pm1}\sum_{\nu=\pm2}
\,\left[\,U_{\mu}(2x_b)U_{\nu}(2x_b+\hat{\mu})+
      U_{\nu}(2x_b)U_{\mu}(2x_b+\hat{\nu})\,\right]
\delta_{x,\,2x_b+\hat{\mu}+\hat{\nu}}\right\}\;\;,
\eea

\noi
where, as usual, $U_{-1}(x) = U^{\dagger}(x-\hat{1})$ and the analogous
relation for $1\rightarrow 2$.

Once the gauge sector has been worked out, with a prescription for
$U^{\rm\pst min}(U)$, we are in a position to solve,
for a given background lattice configuration
$U$, the iteration equation for the FP Dirac operator,
Eq.~(\ref{ffp}). Of course, due to the complexity of the equation,
this cannot be done analytically, and so we have to turn to some
approximate method.
In the next Section we describe the technical details of the numerical method
and discuss the reliability of the approximations.

\section{\bf The numerical computation}
\label{sec:numcom}

The ideas of Sec.~\ref{sec:fpdo} would be of only academic
relevance if we were not able to compute an approximation
of the FP Dirac operator $\Delta^{\rm \pst FP}$ in terms of a matrix
with a finite (and sufficiently short) interaction range.
This approximation should be reliable, in the sense that all the theoretical
properties mentioned in Sec.~\ref{sec:fpdo} should be preserved with good
accuracy. We shall show in the remaining of the paper how this can be done in
practice.

We have computed the FP Dirac operator (or better, some
ultra-local approximation to it, see below) for several
background gauge configurations.
We have considered  the instanton configurations of Eq.~(\ref{inst}) for
$Q=1$ and $Q=2$; we will refer to them as to {\em smooth} configurations.
We have also studied the effect of the super-imposition on these
smooth configurations of a random perturbations with a given size.
This has been done by adding to the gauge field of the smooth
configuration a perturbation of the form
$\delta A_\mu={\rm size}\,\times\, r$, where $r$ is a random number between
$-1$ and $1$. Finally, we have also considered completely
random configurations.

We solve the FP equation (\ref{rra}) by iterations.
Our strategy is the following: we start from some lattice definition of the
Dirac operator and, for a fixed background configuration $U$,
we let it evolve under RG according to Eq.~(\ref{ffp}).
After a large enough number of RG transformations, the running
Dirac operator will converge with sufficient accuracy to the
FP Dirac operator.

In order to obtain, according to Eq.~(\ref{ffp}), the Dirac operator
evolved through $n$ RG transformations $\Delta^{\pss (n)}(U)$,
defined on a $N\times N$ toroidal lattice,
we proceed in the following way: first of all we construct by
iterative procedure the sequence of gauge field configurations:
$U^{\pst(0)}(U)\equiv U, U^{\pst(1)}(U), \ldots, U^{\pst(n)}(U)$,
living respectively on a
$N\times N$, $2 N\times 2 N, \ldots,2^{n}N\times 2^{n}N$ toroidal lattice.
Eq.~(\ref{ffp}) allows us to calculate
$\Delta^{\pss (1)}(U^{\pst(n-1)}(U))$, defined on a $2^{n-1}N\times 2^{n-1}N$
lattice, by putting $\Delta^{\pss (0)}(U^{\pst(n)}(U))$,
defined on a $2^{n}N\times 2^{n}N$ lattice, on the r.h.s.
of the equation; once obtained
$\Delta^{\pss (1)}(U^{\pst(n-1)}(U))$,
the procedure can be iterated to determine
$\Delta^{\pss (2)}(U^{\pst(n-2)}(U))$
... (defined on smaller and smaller lattices)
up to $\Delta^{\pss (n)}(U)$ on the final $N \times N$ lattice.

Considering that the dimension of the matrices entering the above
described procedure explodes when increasing $n$ (taking into
account both space-time and Dirac indices, the largest matrix
involved, $\Delta^{\pss (0)}(U^{\pst(n)}(U))$,
has dimension $2^{2n+1}N^2$) it is clear that
some approximation must be introduced\footnote{Note that in Eq.~(\ref{ffp})
a matrix inversion is required.}.
In what follows we discuss these approximations and the convergence of the
iterative procedure.

\subsection{Approximations \label{sub:appr}}

From RG theoretical arguments, we expect the FP Dirac operator to be local,
though coupling fermions at arbitrarily large
distances. In this context locality means that the matrix elements
$\Delta^{\rm \pss FP}_{x y}$ are exponentially suppressed with
the distance $|x-y|$. We know from a perturbative
computation \cite{fv} that the most important couplings
are those inside the region $|x_\mu-y_\mu|\;\leq\; 2$.
This local structure is very important from the practical point of
view, since it allows us to escape the memory-space problems connected
with the allocation of matrix elements of the Dirac operator
on the finer lattices.

Therefore, as an approximation we consider only ultra-local
fermion matrices: we neglect in all stages of the recursive procedure
the matrix elements which couple fermions at a distance larger than
a certain maximum range $r_g$. In other words, we make a sharp cut in
the interaction, considering a fermion action of the form:

\be
\sum_{\xb,\,l}\,\bar\psi_{\xb}\: \Delta_{\xb\,\xb+l}\:\:\psi_{\xb+l}\;\;,
\ee

\noi
where $l_\mu$, $\mu=1,2$, takes values between $-r_g$ and $r_g$.
It is worthwile to stress that this ultra-local approximation maintains
all the symmetries of the exact action, including hermiticity (Eq.(~\ref{hc})).
To study the locality of the FP Dirac operator and to keep
the effects of the cut under control, we consider the two cases
$r_g=1$ and $r_g=2$.

Even though within the ultra-local approximation memory problems
are partially solved, we have still to cope with the CPU time,
which grows with the number of iterations as  $\sim 2^n$.
As a consequence, we are not
able to make many iterations, and in practice we have to stop at $n=4$
or $n=5$. This means that we do not get the {\em exact} FP Dirac operator,
but some asymptotic approximation to it.

At the level of free fermions, the sharp cut of the couplings has the effect,
besides an unphysical change of the normalization of the fermion fields,
of generating a fermion mass, with the consequence of bringing the action
outside the critical surface. Being the mass a relevant parameter,
the convergence towards the critical FP is therefore spoiled.

Explicitly, writing the free Dirac operator as

\be
\Delta^{\rm\pss free}_{x y}\;=\;
\sum_\mu\,\rho_\mu(x-y)\,\gamma_\mu\:+\:\lambda(x-y)\;\;,
\ee

\noi
we have the following normalization conditions:

\be
\sum_{x}\,x_\mu\,\rho_\mu(x)\;=\;1\;\;,\;\;\;\;\;\;\;\;\;\;\;\;
\sum_{x}\,\lambda(x)\;=\;0\;\;; \label{c}
\ee

\noi
the first equation sets the normalization of the fermion field, and the second
is the massless condition. To preserve conditions~(\ref{c})
in the course of our approximated iterative procedure, we must add
to each step a multiplicative renormalization of the fields and an additive
renormalization of the mass.

In the interacting case, we use the renormalization constants computed in the
free case. This ensures that the iterative procedure keeps the fermion
action in the universality class of the massless Schwinger model.

The renormalization constants are a measure of the goodness of the
ultra-local approximation: the closer are the field and mass
renormalization respectively to one and zero, the better the
approximation. Table~\ref{tab:rc} displays the renormalizations
as a function of the number of iterations for the cases under study
$r_g=1$ and $r_g=2$. We see that for $r_g=2$ the renormalizations are
very small ($\,O(10^{-3})\,$), as expected from the locality properties
of the FP Dirac operator.

\begin{table}
\centering
\begin{tabular}{cccccc}
 & \multicolumn{2}{c}{$r_g=1$} & \hspace{4mm} & \multicolumn{2}{c}{$r_g=2$} \\
\cline{2 - 3} \cline{5 - 6}
\multicolumn{6}{c}{\vspace{-3mm} } \\
step \hspace{2mm} & field & mass & & field & mass \\
\multicolumn{6}{c}{\vspace{-3mm} } \\
1 \hspace{2mm} & 1.1957 & $-0.7431$ & & 1.0026 & \ \ 0.0038 \\
2 \hspace{2mm} & 1.1661 & $-0.7467$ & & 0.9998 & \ \ 0.0040 \\
3 \hspace{2mm} & 1.1508 & $-0.7477$ & & 0.9999 & \ \ 0.0022 \\
4 \hspace{2mm} & 1.1428 & $-0.7482$ & & 1.0019 & \ \ 0.0002 \\
5 \hspace{2mm} & 1.1381 & $-0.7491$ & & 1.0018 & \hspace{-2 truemm} $-0.0020$
\\
\end{tabular}
\caption{The values of the field and mass renormalization constants
after each RG step, for the two sharp cuts considered.}
\label{tab:rc}
\vspace{0.2cm}
\end{table}

\subsection{Convergence}

We call $\Delta^{\pss (0)}(U)$ the starting discretization of the Dirac
operator. In order to improve the convergence of the RG iteration,
it is important to choose such operator as close as possible to the FP.
For the construction of an approximate expression of the
FP Dirac operator, the analytical knowledge of the FP matrix
for the free fermions, $\Delta^{\rm\pst FP\,free}_{x y}$, has been
exploited. The idea is to transform the couplings between two free fields,
$\bar{\psi}_x$ and  $\psi_y$, into gauge-covariant
couplings between fermions interacting with a gauge field;
the simplest way of doing this consists in multiplying
 $\Delta^{\rm\pst FP\,free}_{x y}$
by the lattice parallel transporter on the shortest
path\footnote{In case of more than one
path, an average that keeps the lattice symmetries has been taken.}
from $x$ to $y$. The result is our definition of
$\Delta^{\pss (0)}(U)_{x y}$.
In order to avoid complicated paths, we have restricted the couplings of
$\Delta^{\pss (0)}(U)_{x,x^{\prime}}$ inside the region
$|x_\mu-x^\prime_\mu|\leq1$.

As a criterion to study the convergence of the iterative process, we have
chosen the operator norm:

\be
\| A \|_{\infty}\;=\;
\sup_{\{v\}}\,\left\{\,\frac{\| A\,v\|_2}{\|v\|_2}\,\right\}\;\;,
\ee

\noi
where $v$ is any vector and $\| v\|_2 = \sqrt{\sum v_i^*\,v_i}$. In
Table~\ref{tab:conv} we show the values of the norm of the difference
of two Dirac operators, corresponding to two consecutive iterations,
in the free case and, in the interacting case, for the smooth
background configuration given by Eq.~(\ref{inst}),
random fluctuations around it (sizes 0.5 and 1), and a completely
random configuration; the background configurations
have all fixed point topological charge $Q^{\rm \pss FP}=1$.
The lattice size is $N=6$ (except for the case of the random configuration,
where $N=4$) and the range of the interaction $r_g=1$.
In all the other cases (different lattice size, range $r_g$ and topological
charge) the results are qualitatively similar. Quantitatively, the
deviations from the numbers of Table~\ref{tab:conv} are not large,
$\sim 10\%$ at most. Observe that the data of Table~\ref{tab:conv}
indicate, as expected, the presence of irrelevant operators of dimension three.

We have observed that the rougher the configuration is, the worst the
convergence (the numbers of Table~\ref{tab:conv} are an example).
This is natural, since for a rough configuration $U$ more
iterations are required in order that the configuration at the finest
level, $U^{\pst(n)}(U)$, becomes smooth enough to ensure convergence.
This explains in particular the quite slow convergence for completely random
configurations, as is manifest in Table~\ref{tab:conv}.

\begin{table}
\centering
\begin{tabular}{cccccc}
$i$ & free & smooth & pert. 0.5 & pert. 1 & random \\
\multicolumn{4}{c}{\vspace{-3mm}} \\
\cline{1 - 6}
\multicolumn{4}{c}{\vspace{-3mm}} \\
1 & 0.3320 & 0.3248 & 0.4138 & 0.5989 & 1.9678 \\
2 & 0.1997 & 0.1971 & 0.2068 & 0.2276 & 1.6476 \\
3 & 0.1148 & 0.1143 & 0.1184 & 0.1284 & 1.4608 \\
4 & 0.0630 & 0.0631 & 0.0661 & 0.0740 & 0.7542 \\
5 & 0.0348 & 0.0352 & 0.0382 & 0.0461 & 0.4112 \\
\end{tabular}
\caption{The value of  $\|\Delta^{\pss (i)}-\Delta^{\pss (i-1)}\|_\infty$
for two consecutive iterations, $i-1$ and $i$
($i=0$ is the starting operator).
The case considered is $N=6$, $r_g=1$ and $Q^{\rm\pss FP}=1$.
For the random configuration, $N=4$.}
\label{tab:conv}
\vspace{0.2cm}
\end{table}

\section{\bf Numerical results}
\label{sec:numres}

In this Section we discuss the numerical results coming from the
checks of the classical properties of the FP Dirac operator.

In the iteration of the FP equation a
matrix inversion must be performed at each step. We use for this
an recursive algorithm, which gives results with a precision of $10^{-4}$.
This means that in the following, numbers smaller $10^{-4}$
are compatible with zero.

\subsection{Spectrum and bounds}

We start by discussing the properties of the spectrum.
We use $\kappa_{\rm \pst F}=4$, since in the free case this
value gives the most local action. Eq.~(\ref{bound}) for $q=1/2$ tells us that
the spectrum must be inside the circle of the complex plane of radius
$2$ centered at $(2,0)$ and outside the circle of radius $1$
centered at $(1,0)$. While in no case we have found a violation of the
first bound (it is true for each value of the number of iterations,
in fact, it turns to be true also for the starting operator
$\Delta^{\pss (0)}$), the second one holds with good approximation
only for $r_g=2$.

\begin{figure}[t]
\centering
\begin{minipage}[t]{5 cm}
\setlength{\unitlength}{1cm}
\mbox{\psfig{file=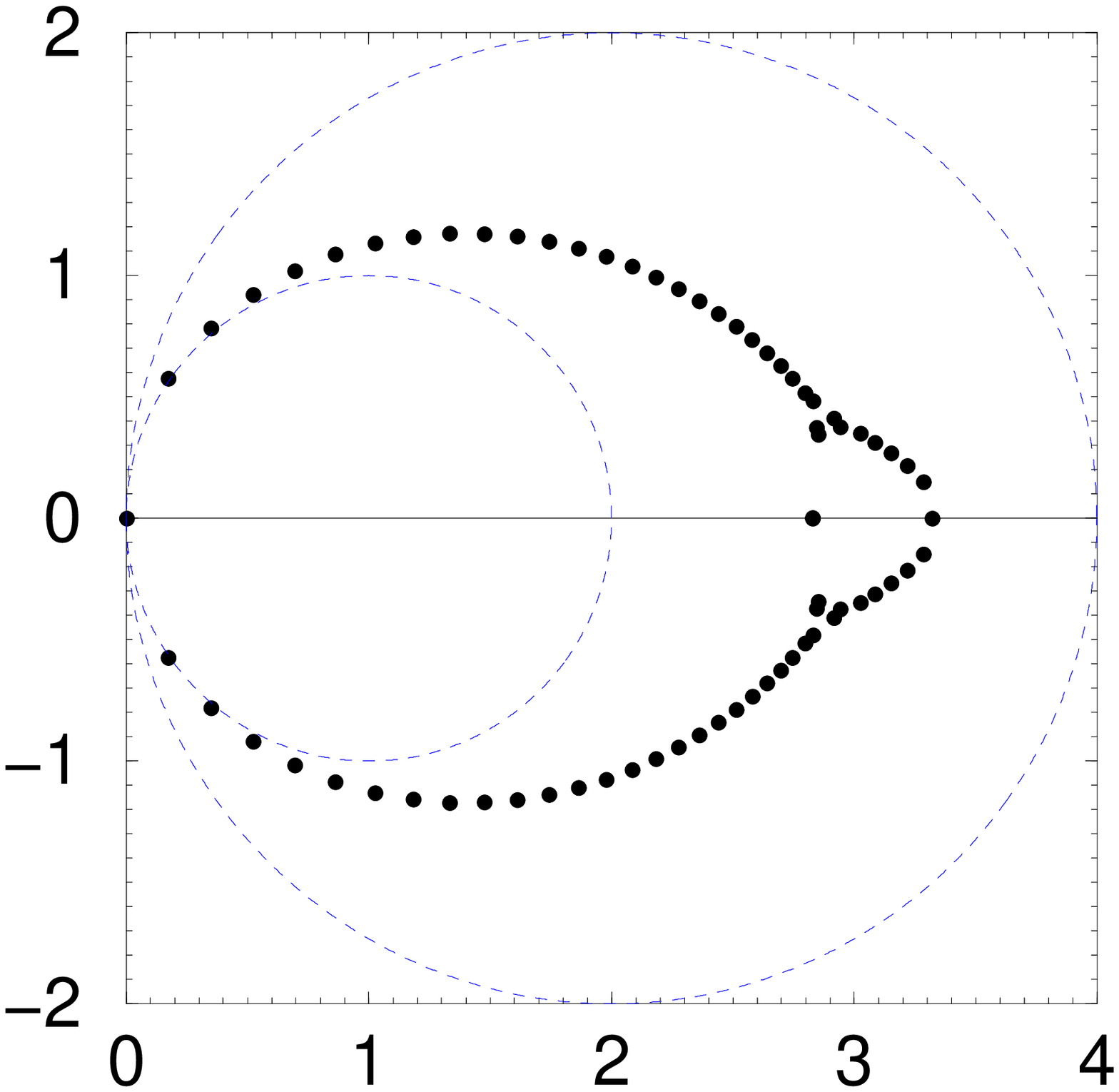,width=5 truecm, angle=-90}\put(-1,-4.3){\bf a}}
\mbox{\psfig{file=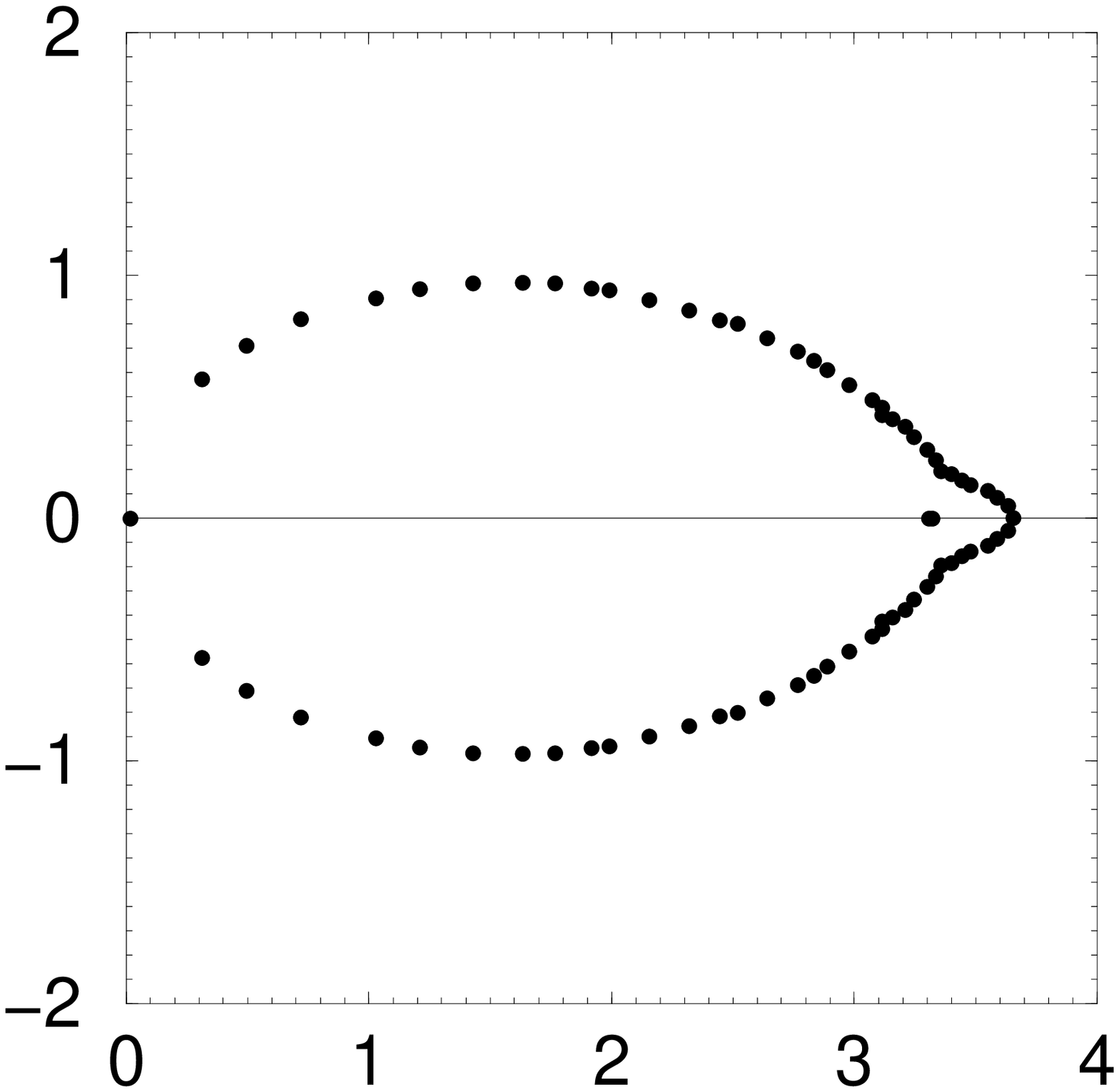,width=5 truecm, angle=-90}\put(-1,-4.3){\bf c}}
\end{minipage}
\begin{minipage}[t]{5 cm}
\setlength{\unitlength}{1cm}
\mbox{\psfig{file=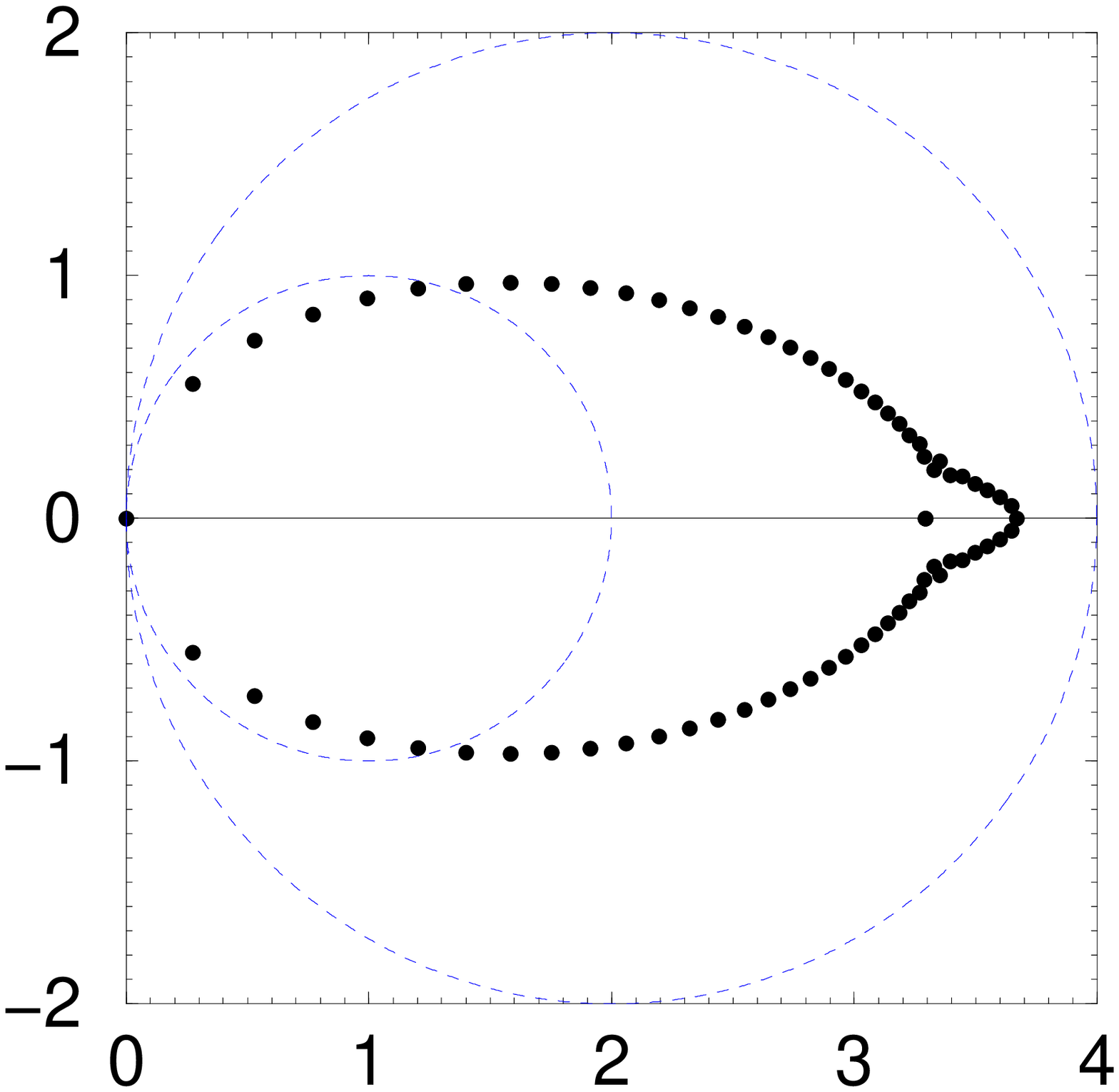,width=5 truecm, angle=-90}\put(-1,-4.3){\bf b}}
\mbox{\psfig{file=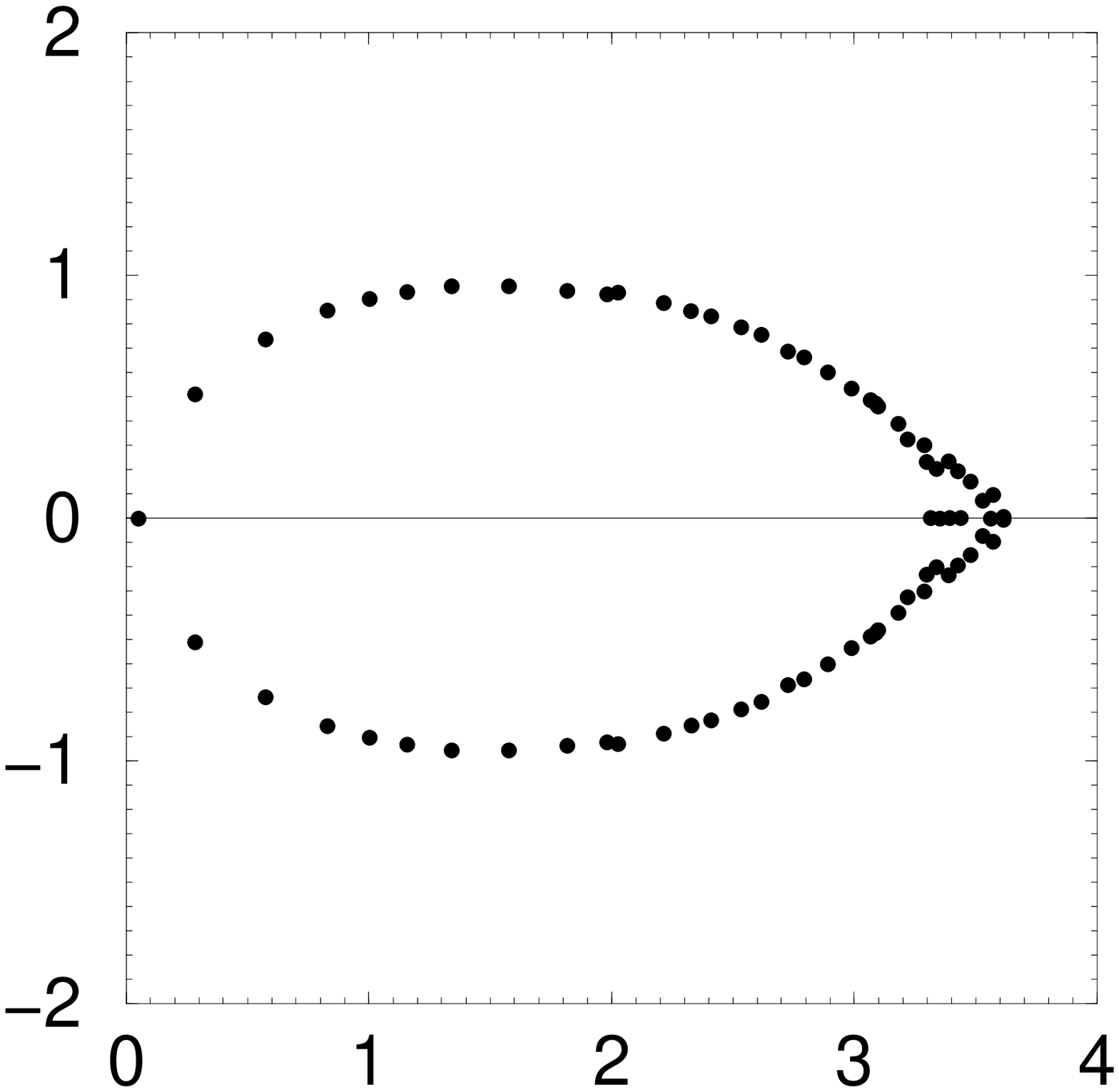,width=5 truecm, angle=-90}\put(-1,-4.3){\bf d}}
\end{minipage}
\caption{The spectrum of the FP Dirac operator $\Delta^{\rm\pst FP}$.
The lattice size is
$N=6$ and the number of iterations $n=5$; a)
smooth background configuration with $Q^{\rm \pst FP} =1$; the range of the
couplings considered in the action is $r_g=2$ (see subsection \ref{sub:appr});
the two circles represent the bounds on the spectrum, as explained in the text;
b) as in a), but with $r_g=1$; c) and d)
$r_g$=1 and background configuration with perturbation of size 0.5 and
1 respectively (still $Q^{\rm \pst FP}=1$).}
\label{fig:spe}
\vspace{0.5cm}
\end{figure}

Fig.~\ref{fig:spe} displays the spectrum of the FP Dirac operator
on a $6^2$ lattice for background configurations with $Q^{\rm \pst FP}=1$.
The Dirac operator has been obtained through 5 RG iterations.
The deformation of the spectrum induced by the cut can be observed
by comparing Figs.~\ref{fig:spe}-a and~\ref{fig:spe}-b,
which refer to the two different cuts $r_g=2$ and $r_g=1$ respectively,
in the case of a smooth background configuration.
We see that for $r_g=2$ the bounds on the eigenvalues are well
satisfied, while for $r_g=1$ a slight violation comes into play
in the left edge of the spectrum due to the approximation
of the sharp cut of the couplings.
Comparison between Figs.~\ref{fig:spe}-b and~\ref{fig:spe}-c,d
allows to evaluate the stability of the spectrum against perturbation
of the background configuration, corresponding the two latter cases
to a perturbation of the background configuration of size 0.5 and 1
respectively (the cut is always $r_g=1$).
In all cases, except d, we find three real eigenvalues
$\lambda_{\pst 0}<\lambda_{\pst 1}<\lambda_{\pst 2}$: the lowest one,
$\lambda_{\pst 0}$ corresponds
to the would be zero mode, see below; $\lambda_{\pst 1}$ is twofold
degenerate. In the case d,
where a strong perturbation of the background configuration is applied,
the real eigenvalues are six, with no degeneracy.
The higher real eigenvalues, lying on the right edge of the spectrum,
are always decoupled from the infrared part of the spectrum, where the
lowest real eigenvalue $\lambda_{\pst 0}$ is located.

\begin{figure}[t]
\centering
\mbox{\psfig{file=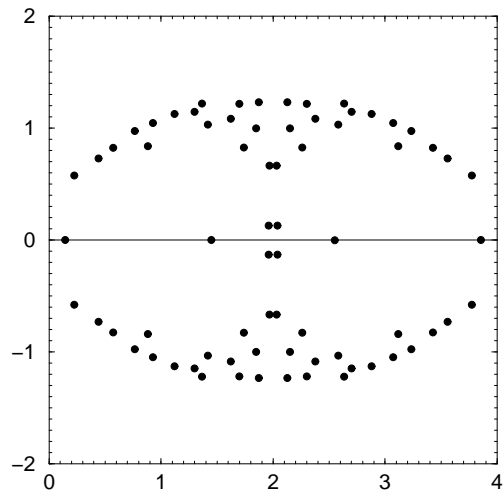,width=8 truecm, angle=-90}}
\caption{The spectrum of the Wilson operator $\Delta^{\pst\rm W}$
on a $6^2$ lattice for the smooth background configuration with
$Q^{\rm \pst FP}=1$.}
\label{fig:wil}
\vspace{0.5cm}
\end{figure}

It is interesting the comparison with the case of the (massless)
Wilson operator, which has been studied recently in Ref.~\cite{ghl};
the eigenvalues fulfil in this case an additional symmetry~\cite{ghl},
the reflection with respect
to the point (2,\,0) in the complex plane\footnote{Actually,
this additional symmetry holds only for even lattices.}.
For comparison with the FP case, we report in Fig.~\ref{fig:wil}
the spectrum of the Wilson operator for the smooth configuration
with topological charge one.

\begin{figure}[t]
\centering
\begin{minipage}[t]{5 cm}
\setlength{\unitlength}{1cm}
\mbox{\psfig{file=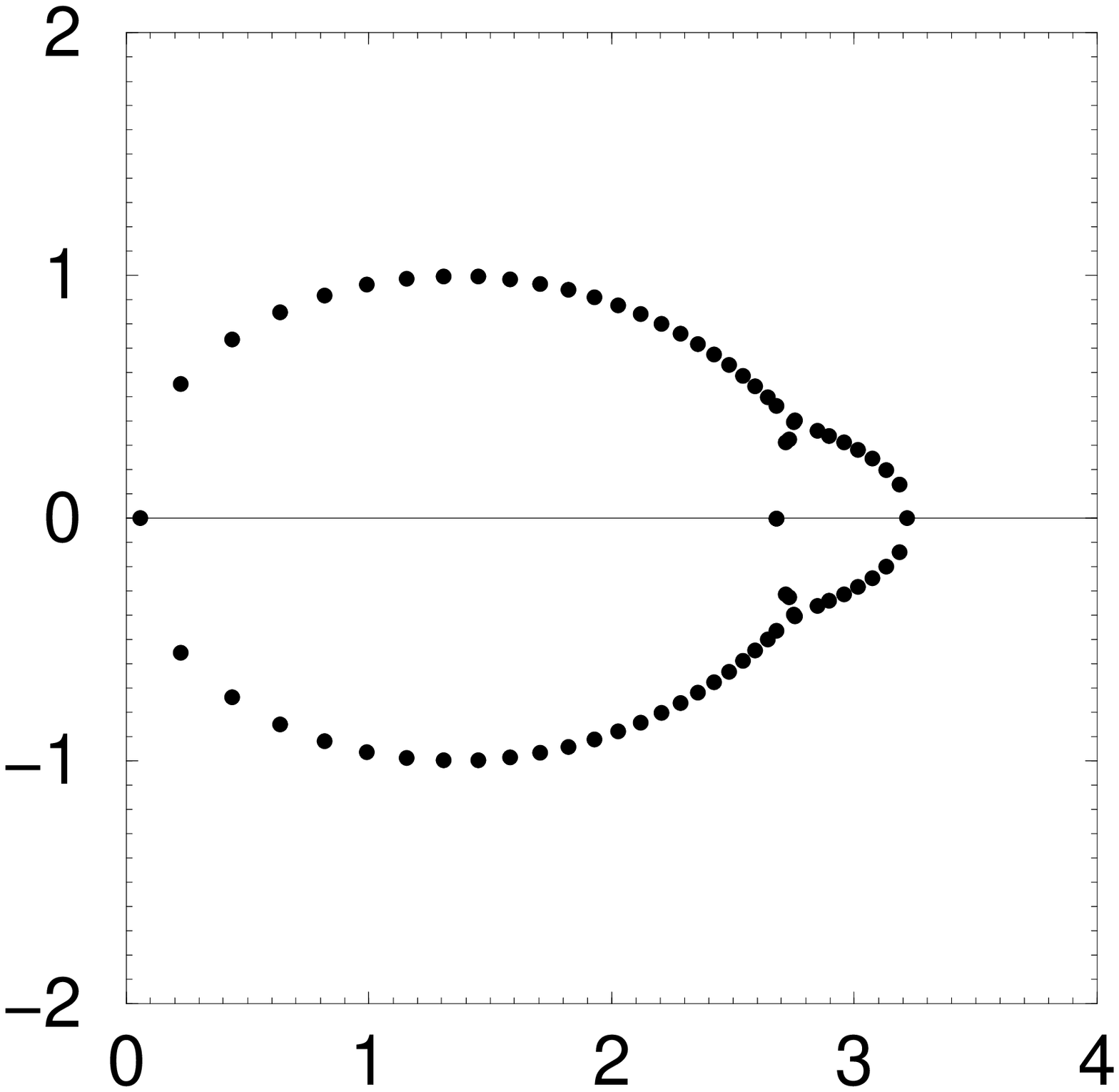,width=5 truecm, angle=-90}\put(-1,-4.3){\bf a}}
\mbox{\psfig{file=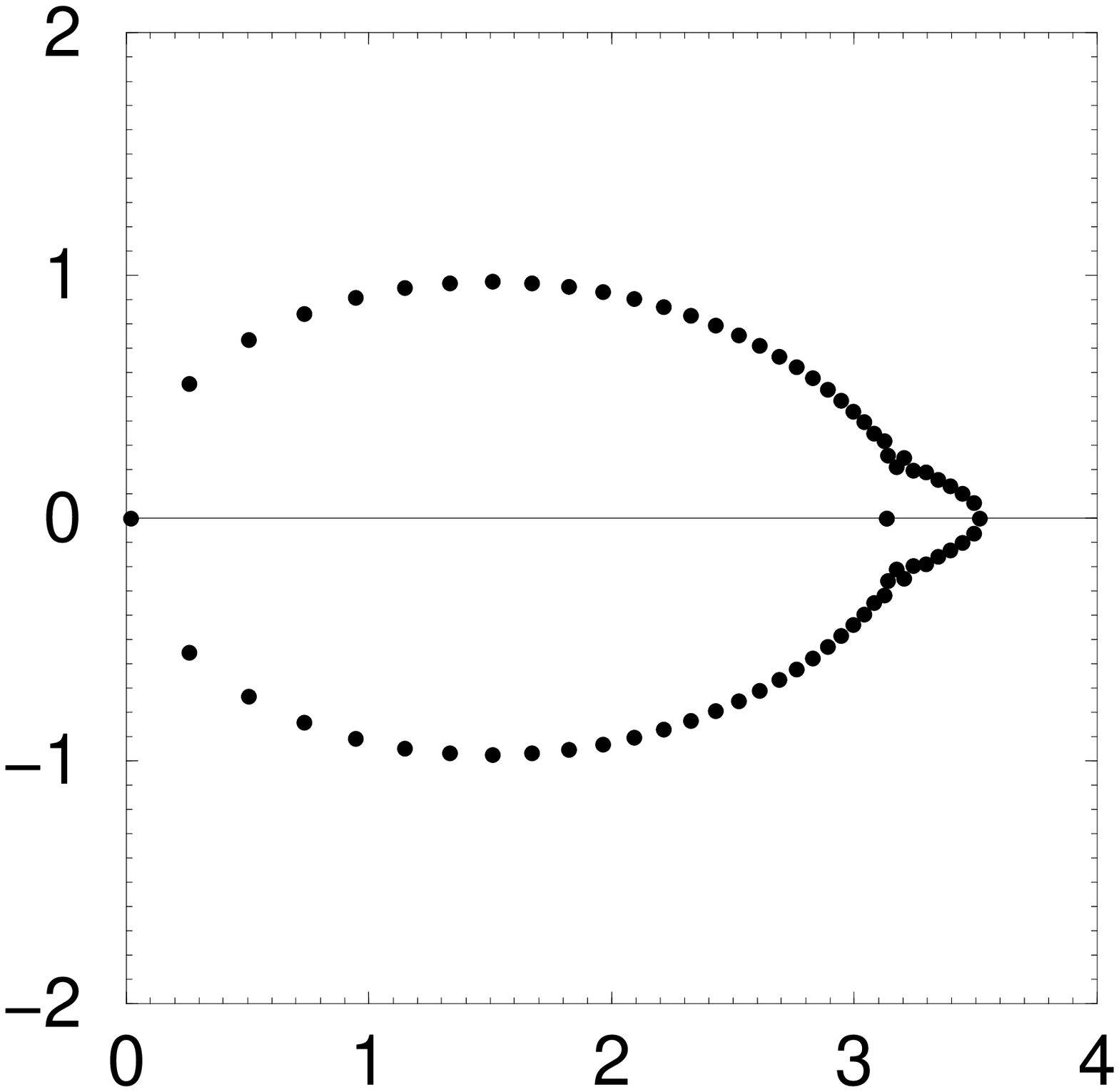,width=5 truecm, angle=-90}\put(-1,-4.3){\bf c}}
\end{minipage}
\begin{minipage}[t]{5 cm}
\setlength{\unitlength}{1cm}
\mbox{\psfig{file=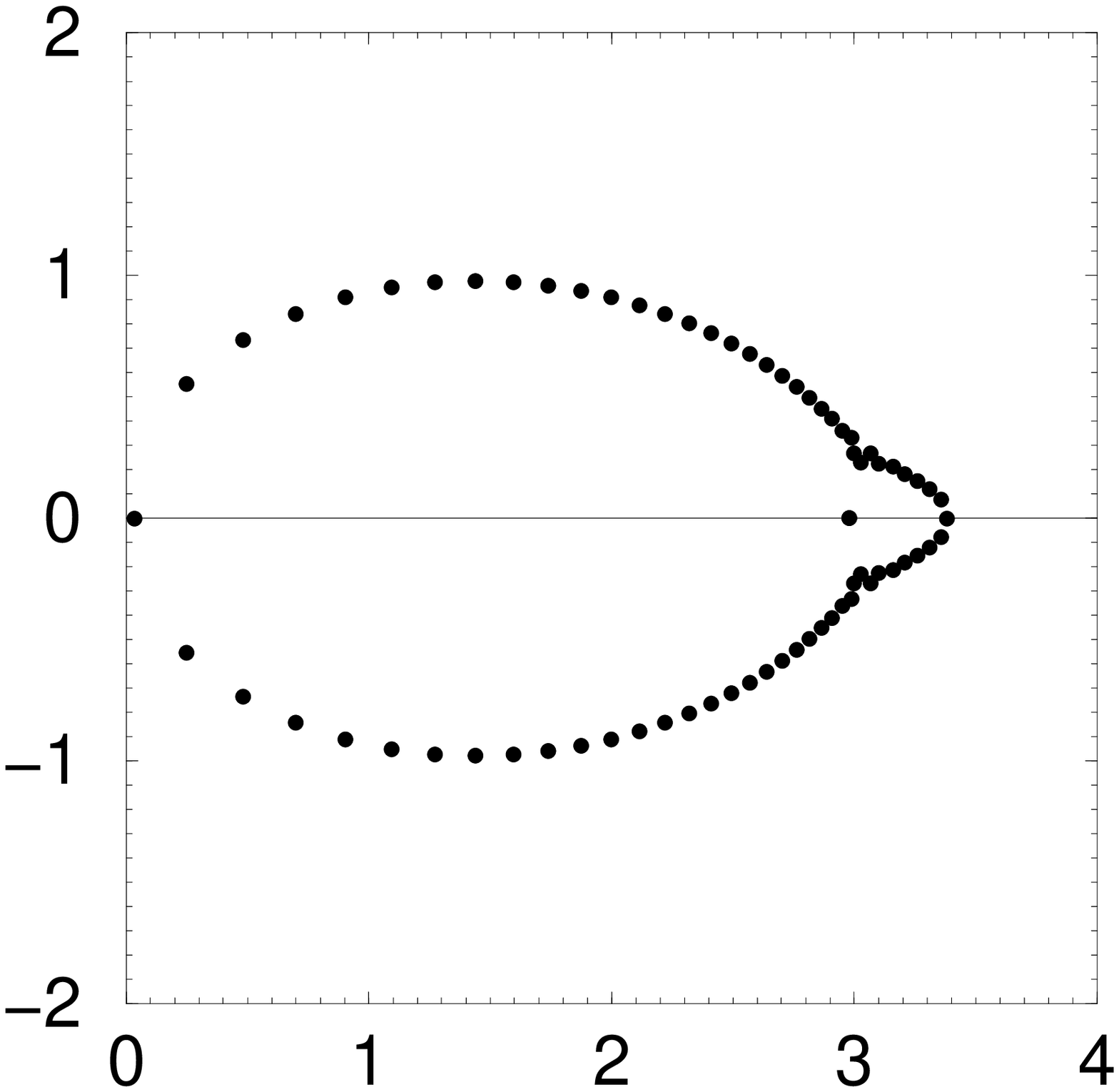,width=5 truecm, angle=-90}\put(-1,-4.3){\bf b}}
\mbox{\psfig{file=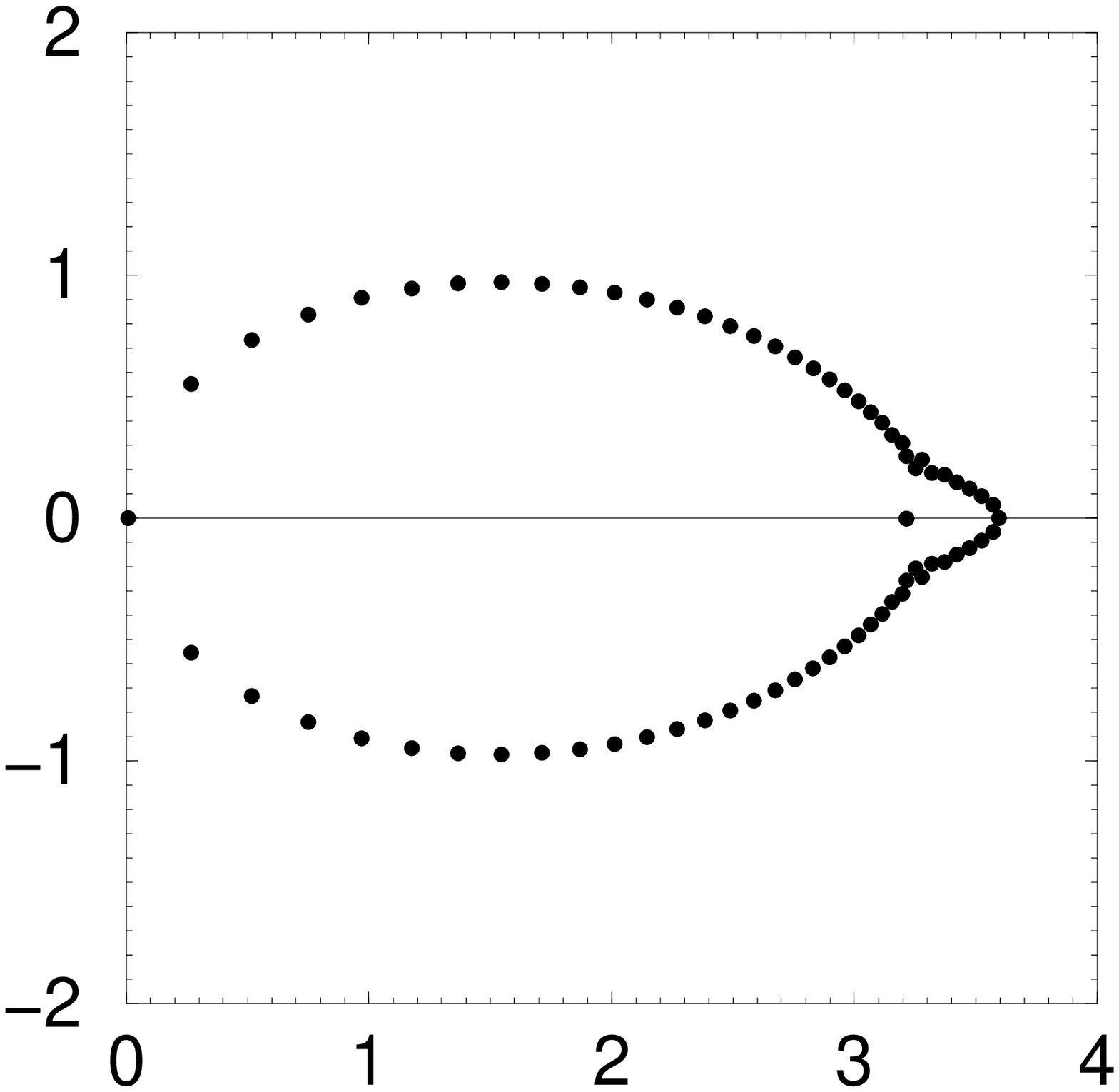,width=5 truecm, angle=-90}\put(-1,-4.3){\bf d}}
\end{minipage}
\caption{The evolution under RG of the spectrum of the Dirac operator
on a $6^2$ lattice starting from  the Wilson discretization
$\Delta^{\pst \rm W}$,
for number of iterations $n=1$~(a), $n=2$~(b), $n=3$~(c), $n=4$~(d).
The smooth background configuration has charge $Q^{\rm \pst FP}=1$
and $r_g=1$.}
\label{fig:evol}
\vspace{0.5cm}
\end{figure}

In Fig.~\ref{fig:evol} the evolution of the spectrum under
the RG process starting from the Wilson operator is displayed,
for $n=1,2,3$ and 4, in the case of the smooth configuration.
We see that already for $n=1$ the spectrum has assumed
the main features of its characteristic geometry, and only small
changes take place thereafter; in particular,
for $n=4$ the convergence to the final
spectrum seems to be reached with good accuracy.
It is also possible to observe
the progressive decoupling of the two  higher real eigenvalues
$\lambda_{\pst 1}$ and $\lambda_{\pst 2}$, drifting  towards higher values,
from the lowest one $\lambda_{\pst 0}$, converging to zero.

\subsection{Zero modes and index theorem}

From the discussion of Sec.~\ref{sec:fpdo} we know that the FP
Dirac operator must have zero modes in the topologically
non trivial sectors, with definite chirality
and satisfying the Index Theorem, Eq.~(\ref{it}), if the continuum
topological charge is replaced by the FP topological charge $Q^{\rm \pst FP}$.
In the Schwinger model the Vanishing Theorem~(\ref{eq:van}) must also be true.
As a consequence, in the case $Q^{\rm \pst FP}>0$, the FP Dirac operator
should have $Q^{\rm \pst FP}$ zero modes with positive chirality.
Since we do not attain exactly the
FP, we do not expect {\em exact} zero modes.
Rather, we expect that the $Q^{\rm \pst FP}$ lowest
lying eigenvalues converge to zero when the number of iterations tends
to infinity. Still, the cut of the couplings in the region
$|x^\prime-x|> r_g$ can produce unwanted effects.

We start the discussion with the smooth configuration.
In Table~\ref{tab:chirals} we can see the eigenvalues of smallest modulo as a
function of the number of iterations~$i$.
We report also the overall chirality of the zero modes (trace of $\gamma_5$
in the subspace of zero modes, $\sum_v (v,\gamma_5\,v)$),
which should equal the FP topological charge of the background configuration.
Notice that the eigenvalues are (within machine precision) real
for $i=0$; this is in agreement
with the observation~\cite{ghl} that eigenvectors with non-zero average
chirality must have necessarily real eigenvalues.
This is an {\em exact} statement, relying only on the hermiticity property
of the fermion matrix~(\ref{hc}), which is preserved by the RG procedure.
The small numbers $O(10^{-6})$
found for the imaginary part of the eigenvalues for $i\ge1$
are due to the approximation introduced
when operating the matrix inversion in the recursion
(the ultra-local approximation does not violate the hermiticity property).
In accordance with the Index and Vanishing Theorem, in the case
$Q^{\rm \pst FP}=1$ we find only one non-degenerate small real eigenvalue,
while for $Q^{\rm \pst FP}=2$ the lowest lying eigenvalue
is twofold degenerate. The real part of the eigenvalues converges
towards zero when increasing the number of iterations;
the law is roughly as $1/2^i$, as expected
from the dimensional analysis of the lowest dimensional irrelevant
operators. This law breaks down for $i=5$ in the case $r_g=1$,
where we find numbers smaller than expected.
The eigenvectors have in any case definite positive chirality,
regardless to the number of iterations.
In fact, this is already true for the starting
Dirac operator, and the cut of the couplings seems not to
be effective in this regard.

In conclusion, our results are consistent with the picture
of a FP Dirac operator {\em exactly} verifying the Index Theorem, and
being well approximated by ultra-local fermion matrices.

It is interesting to compare these results with those given by the Wilson
operator, displayed in Table~\ref{tab:chiralw}. Notice how the
Index theorem is only asymptotically verified for
$N\rightarrow\infty$, i.e., in the continuum limit.

The numbers of Table~\ref{tab:chiralf} have been obtained by adding a random
perturbation of size $0.5$ to the smooth configurations
previously considered (see also Table~\ref{tab:chiralf1}).
Notice that the lowest lying eigenvalues are
about twice larger than the corresponding for the smooth
configuration. Again they decrease by a factor of about 2 after each
iteration. Notice also that the random perturbation removes
the degeneracy  of the two lowest lying eigenvalues in the case
$Q^{\rm \pst FP}=2$.
Finally, we see that the chirality is no more perfect from the
start, but already for $i=4$ it is compatible with 1 within
our numerical accuracy.

\subsection{The topological charge}

The FP topological charge can be obtained directly from the FP Dirac
operator through the formula~(\ref{top}).
This relation could be useful for the calculations of topological
quantities in unquenched QCD with the FP action, since it gives automatically
a prescription for the FP topological charge, once a parametrization of the FP
fermion matrix has been found.
In our approach, we can check
how accurately this relation is verified as the number of iterations
of the FP equation is increased.
The results are displayed in Table~\ref{tab:topgw}.
We see that the difference with the FP topological charge is less
than $14\%$ after $4$ iterations and less than $4\%$ after $5$ iterations
with $r_g=1$ in the case $Q=1$. Moreover, the results are stable under
the perturbation of the background configuration.
With $r_g=2$ we observed only a slight improvement.

\section{\bf Conclusions}

We have presented an explicit example which illustrates that it is possible
to compute an ultra-local approximation of the FP Dirac operator
which preserves up to a very good accuracy all the important properties
of the exact FP operator. We showed
by numerical computation that in the case of the Schwinger model
it is a very good approximation to consider only couplings
between fermions contained in the same size-two plaquette
(translated in 4 dimensions: size-two hypercube).
We have confined our work to the classical properties, like zero
modes and their chirality, topological charge and the Index Theorem.
We have checked these properties in the case of instanton solutions
of the equation of motion and random perturbations around them.
We believe, however, that the quantum properties of the FP action, like
for example the
non-renormalization of the pion mass, will also hold with similar accuracy.
An indication of this is that another property derived from the GWR,
namely the relation~(\ref{top}) between the FP Dirac operator
and the topological charge of the background gauge configuration,
holds within this same accuracy also when introducing quantum fluctuations
around the classical instanton solutions. We are nevertheless conscious
of the fact~\cite{ghl2} that these mildly perturbed configurations are
not representative of the whole MC set of configurations
at thermal equilibrium.

Recently a great attention has been paid to the Index theorem on the lattice.
The goal was to find some traces of it, either using the standard Wilson
action~\cite{ghl,ghl2}, some chiral improved version of it~\cite{pilar} or
approximated solutions of the GW relation constructed from the overlap
formalism~\cite{nara}. In all these cases, the Index theorem seems to hold
only in a probabilistic sense. We hope to have convinced the reader that
the exact Index theorem of Ref.~\cite{hln} holds very accurately
even with an ultra-local approximation of the FP Dirac operator.

Our ultimate conclusion is therefore that a parametrization
of the FP action, involving only very localized couplings,
with good scaling properties should be possible.
Moreover, following the suggestion by Hasenfratz and
Niedermayer, one can use the freedom typically present in this kind of
fits to enforce up to a very high precision the classical properties
for some instanton-like smooth configurations.
A parametrization of the FP action of the Schwinger model was
already found in~\cite{graz}, but locality was not the first priority
of the authors in that work. Really, this is urgent problem only in
view of four dimensional theories.

We solved the FP equation by iterations. The analytical
solution of the pure gauge part (this is an extra-bonus of the
two dimensions) enabled us to concentrate the numerical
effort in the fermion sector. Also the low dimensionality of the model was of
great help, but still, it was not possible to make many iterations.
With five we got a good precision. In four dimensions, however,
memory-space and CPU-time problems are likely to be an obstacle to
this naive iterative approach. Different strategies to solve the FP equation
should be devised. A possibility to avoid extremely large lattices is to
use, as in~\cite{graz}, a parametrization of the FP action
from the very beginning, namely to solve the iteration equation
in a finite space of couplings.

\vspace{1cm}

\noindent
{\bf Acknowledgments.}

We are indebted with P.~Hasenfratz and F.~Niedermayer
for stimulating discussions and useful suggestions.
The work received financial support from
Fonds zur F\"orderung der Wissenschaftlichen Forschung
in \"Osterreich - Austria, project P11502-PHY (F.F.), and Ministerio
de Educaci\'on y Cultura - Spain under grant PF-95-73193582 (V.L.).

\newpage

\newpage

\newpage

\newpage

\begin{table}[t]
\centering
\begin{tabular}{cccccccc}
\multicolumn{8}{l}{$Q = 1$} \\
 & \multicolumn{3}{c}{$r_g=1$} & \hspace*{6mm}
& \multicolumn{3}{c}{$r_g=2$} \\
\multicolumn{8}{c}{\vspace{-2mm}} \\
\cline{2 - 4} \cline{6 - 8}
\multicolumn{8}{c}{\vspace{-1mm}} \\
$i$\hspace*{2mm} & ${\textstyle\rm Re}\,\lambda$ & ${\textstyle\rm
Im}\,\lambda$ &
$\sum_v\,(v,\,\gamma_5 v)$  & \hspace*{6mm} &
${\textstyle\rm Re}\,\lambda$ & ${\textstyle\rm Im}\,\lambda$ &
$\sum_v\,(v,\,\gamma_5 v)$ \\
\multicolumn{8}{c}{\ \ } \\
0\hspace*{2mm} & 0.1240 & \ \ 6.0$\:\:10^{-16}$ & 1.0000 & \hspace*{6mm}
& ---  & ---  & ---    \\
1\hspace*{2mm} & 0.0778 & \ \ 1.4$\:\:10^{-7}$ & 1.0000 & \hspace*{6mm}
& 0.0611 & $-3.7$$\:\:10^{-7}$ & 1.0000 \\
2\hspace*{2mm} & 0.0466 & \ \ 7.3$\:\:10^{-7}$ & 1.0000 & \hspace*{6mm}
& 0.0290 & $-1.2$$\:\:10^{-6}$ & 1.0000 \\
3\hspace*{2mm} & 0.0250 & $-1.3$$\:\:10^{-6}$ & 1.0000 & \hspace*{6mm}
& 0.0134 & $-4.9$$\:\:10^{-6}$ & 1.0000 \\
4\hspace*{2mm} & 0.0101 & \ \ 6.4$\:\:10^{-6}$ & 1.0000  & \hspace*{6mm}
& 0.0065 & \ \ 9.6$\:\:10^{-7}$ & 1.0000 \\
5\hspace*{2mm} & 0.0017 & \ \ $7.4$$\:\:10^{-9}$ & 1.0000 & \hspace*{6mm}
& 0.0039 & $-5.2$$\:\:10^{-6}$  & 1.0000  \\
\multicolumn{8}{c}{\ \ } \\
\multicolumn{8}{l}{$Q = 2$} \\
\multicolumn{8}{c}{\ \ } \\
0\hspace*{2mm} & 0.2441 & $-3.0$$\:\:10^{-16}$ & 2.0000 &\hspace*{6mm}
& ---  & ---  & ---    \\
1\hspace*{2mm} & 0.1570 & \ \ 3.6$\:\:10^{-8}$ & 1.9998 & \hspace*{6mm}
& 0.1233 & $-1.9$$\:\:10^{-6}$ & 2.0000 \\
2\hspace*{2mm} & 0.0960 & \ \ 8.7$\:\:10^{-7}$ & 1.9998 & \hspace*{6mm}
& 0.0598 & $-3.5$$\:\:10^{-6}$ & 2.0000 \\
3\hspace*{2mm} & 0.0531 & $-7.6$$\:\:10^{-7}$ & 1.9998 & \hspace*{6mm}
& 0.0281 & $-6.8$$\:\:10^{-6}$ & 2.0000 \\
4\hspace*{2mm} & 0.0233 & \ \ 6.8$\:\:10^{-6}$ & 1.9998 & \hspace*{6mm}
& 0.0135 & \ \ 1.4$\:\:10^{-6}$ & 2.0000 \\
5\hspace*{2mm} & 0.0026 & $-1.3$$\:\:10^{-5}$ & \ \ \ \ \ \ 2.0000
\hspace*{6mm}
&   &   &    \\
\end{tabular}
\caption{The lowest lying eigenvalue and the chirality of the corresponding
eigenvectors of the Dirac operator for a smooth background gauge
configuration ($Q=1$ and $2$), for each number of iterations $i$. The lattice
size is $N=6$. $r_g$ is the range over which the approximated action extends,
see subsection \ref{sub:appr}.}
\label{tab:chirals}
\vspace{0.2cm}
\end{table}

\begin{table}
\centering
\begin{tabular}{cccc}
$N$ & ${\textstyle\rm Re} \,\lambda$ & ${\textstyle\rm  Im} \,\lambda$ &
$(v,\,\gamma_5 v)$\\
\multicolumn{4}{c}{\vspace{-3mm}} \\
4  & 0.3183  & \ \ 1.2$\:\:10^{-16}$ & 0.8258 \\
6  & 0.1438  & \ \ 1.2$\:\:10^{-15}$ & 0.8372 \\
8  & 0.0805  & \ \ 5.7$\:\:10^{-17}$ & 0.8678 \\
10 & 0.0504  & \ \ 6.9$\:\:10^{-17}$ & 0.8909 \\
12 & 0.0340  & \ \ 2.5$\:\:10^{-16}$ & 0.9073 \\
14 & 0.0242  & $-1.8$$\:\:10^{-16}$ & 0.9193 \\
16 & 0.0180  & $-4.9$$\:\:10^{-16}$ & 0.9284 \\
18 & 0.0138  & \ \ 4.8$\:\:10^{-15}$ & 0.9357 \\
20 & 0.0109  & \ \ 1.4$\:\:10^{-15}$ & 0.9415 \\
22 & 0.0088  & $-5.5$$\:\:10^{-15}$ & 0.9463 \\
24 & 0.0073  & \ \ 3.6$\:\:10^{-15}$ & 0.9505 \\
\end{tabular}
\caption{The lowest lying eigenvalue and the chirality of the corresponding
eigenvector of the Wilson Dirac operator on lattices of different sizes $N$,
with the smooth background configuration having $Q=1$.}
\label{tab:chiralw}
\vspace{0.2cm}
\end{table}

\begin{table}
\centering
\begin{tabular}{cccccccc}
\multicolumn{8}{l}{$Q = 1$} \\
 & \multicolumn{3}{c}{$r_g=1$} & \hspace*{6mm}
& \multicolumn{3}{c}{$r_g=2$} \\
\multicolumn{8}{c}{\vspace*{-2mm}} \\
\cline{2 - 4} \cline{6 - 8}
\multicolumn{8}{c}{\vspace{-1mm}} \\
$i$\hspace*{2mm} & ${\textstyle\rm Re}\,\lambda$ & ${\textstyle\rm
Im}\,\lambda$ &
$(v,\,\gamma_5 v)$  & \ \hspace*{6mm} &
${\textstyle\rm Re}\,\lambda$ & ${\textstyle\rm Im}\,\lambda$ &
$(v,\,\gamma_5 v)$ \\
\multicolumn{8}{c}{\ \ } \\
0 \hspace{2mm} &  0.2109 & $-1.0$$\:\:10^{-15}$  & 0.9739 & \hspace{6mm}
& ---    &  ---  & ---   \\
1 \hspace{2mm} &  0.1242 &  $-1.1$$\:\:10^{-5}$ & 0.9938 & \hspace{6mm}
& 0.0966 & \ \ 5.4$\:\:10^{-7}$ & 0.9938 \\
2 \hspace{2mm} & 0.0797 & $-9.9$$\:\:10^{-6}$ & 0.9980 & \hspace{6mm}
& 0.0450 & \ \ 6.3$\:\:10^{-7}$ & 0.9985 \\
3 \hspace{2mm} & 0.0512 & $-1.1$$\:\:10^{-5}$ & 0.9993 & \hspace{6mm}
& 0.0204 & $-7.1$$\:\:10^{-7}$ & 0.9997 \\
4 \hspace{2mm} & 0.0317 & $-2.7$$\:\:10^{-6}$ & 0.9996 & \hspace{6mm}
& 0.0092 & \ \ 3.6$\:\:10^{-6}$ & 1.0000 \\
5 \hspace{2mm} & 0.0184 & $-2.0$$\:\:10^{-6}$ & 0.9997 & \hspace{6mm}
& 0.0046 & $-2.2$$\:\:10^{-6}$ & 1.0000 \\
\multicolumn{8}{c}{\ \ } \\
\multicolumn{8}{l}{$Q = 2$} \\
\multicolumn{8}{c}{\ \ } \\
0 \hspace{2mm} & 0.2603 & $-4.6$$\:\:10^{-16}$ & 0.9792 & \hspace{6mm}
& --- & --- & --- \\
\vspace{1mm} & 0.3493 & $-8.0$$\:\:10^{-16}$ & 0.9506 & \hspace{6mm}
& --- & --- & --- \\
1 \hspace{2mm} & 0.1589 & \ \ 2.5$\:\:10^{-6}$ & 0.9955 & \hspace{6mm}
& 0.1249 & $-8.2$$\:\:10^{-7}$ & 0.9957 \\
\vspace{1mm} & 0.2150 & $-4.7$$\:\:10^{-6}$ & 0.9884 & \hspace{6mm}
& 0.1676 & \ \ 1.5$\:\:10^{-6}$  & 0.9909 \\
2 \hspace{2mm} & 0.0989 & \ \ 3.7$\:\:10^{-6}$ & 0.9986 & \hspace{6mm}
& 0.0603 & \ \ 7.7$\:\:10^{-7}$ & 0.9989 \\
\vspace{1mm} & 0.1371 & $-8.3$$\:\:10^{-6}$ & 0.9960 & \hspace{6mm}
& 0.0807 & $-1.0$$\:\:10^{-6}$ & 0.9980 \\
3 \hspace{2mm} & 0.0583 & \ \ 6.9$\:\:10^{-7}$ & 0.9995 & \hspace{6mm}
& 0.0377 & $-1.2$$\:\:10^{-6}$ & 0.9996 \\
\vspace{1mm} & 0.0850 & $-9.4$$\:\:10^{-6}$ & 0.9984 & \hspace{6mm}
& 0.0283  & \ \ 4.7$\:\:10^{-6}$ & 0.9998 \\
4 \hspace{2mm} & 0.0299 & \ \ 7.9$\:\:10^{-6}$ & 0.9998 & \hspace{6mm}
& 0.0174 & $-1.5$$\:\:10^{-7}$ & 1.0000 \\
\vspace{1mm} & 0.0490 & $-4.4$$\:\:10^{-6}$ & 0.9992 & \hspace{6mm}
& 0.0133 & \ \ 3.5$\:\:10^{-6}$ & 1.0000 \\
\end{tabular}
\caption{The same as table~\ref{tab:chirals} adding a random fluctuation of
size $0.5$ to the smooth configuration.}
\label{tab:chiralf}
\vspace{0.2cm}
\end{table}

\begin{table}
\centering
\begin{tabular}{cccc}
\multicolumn{4}{c}{\vspace*{-2mm}} \\
\multicolumn{4}{c}{\vspace{-1mm}} \\
$i$\hspace*{2mm} & ${\textstyle\rm Re}\,\lambda$ & ${\textstyle\rm
Im}\,\lambda$ &
$(v,\,\gamma_5 v)$  \\
\cline{1 - 4}
\multicolumn{4}{c}{\vspace{-1mm} } \\
0 \hspace{2mm} & 0.3348 & \ \ 2.0$\:\:10^{-16}$ & 0.9282 \\
1 \hspace{2mm} & 0.1824 & \ \ 1.1$\:\:10^{-5}$  & 0.9869 \\
2 \hspace{2mm} & 0.1245 & \ \ 9.7$\:\:10^{-6}$  & 0.9963 \\
3 \hspace{2mm} & 0.0908 & \ \ 5.7$\:\:10^{-6}$  & 0.9986 \\
4 \hspace{2mm} & 0.0679 & \ \ 8.7$\:\:10^{-6}$  & 0.9990 \\
5 \hspace{2mm} & 0.0520 & $-2.2$$\:\:10^{-6}$ & 0.9989 \\
\end{tabular}
\caption{The same as table~\ref{tab:chirals} adding a random fluctuation of
size $1$ to the smooth configuration. In this case $Q=1$ and $r_g=1$.}
\label{tab:chiralf1}
\vspace{0.2cm}
\end{table}

\begin{table}[t]
\centering
\begin{tabular}{cccccccc}
 & \multicolumn{3}{c}{ $Q=1$ } & \hspace{4mm} &
\multicolumn{3}{c}{ $Q=2$ } \\
\multicolumn{8}{c}{\vspace*{-2mm}} \\
\cline{2 - 4} \cline{6 - 8}
\multicolumn{8}{c}{\vspace{-2mm}} \\
$i$\hspace*{2mm} & smooth & Pert. 0.5 & pert. 1 & & smooth & Pert. 0.5
& Pert. 1 \\
\multicolumn{8}{c}{\vspace{-2mm}} \\
1 & 0.2289 &  0.2280 & 0.2118 & & 0.4565 & 0.4498 & 0.4263 \\
2 & 0.5114 &  0.5067 & 0.4906 & & 1.0067 & 0.9954 & 0.9691 \\
3 & 0.7223 &  0.7185 & 0.7100 & & 1.4143 & 1.4020 & 1.3902 \\
4 & 0.8668 &  0.8637 & 0.8663 & & 1.6959 & 1.6839 & 1.6878 \\
5 & 0.9631 &  0.9608 & 0.9740 & & 1.8844 &  ---   &  ---    \\
\end{tabular}
\caption{The value of $-\f{1}{2\kappa_{\rm \pst F}}\,{\rm tr}\,
\left(\gamma_5\,\{\dfp ,\,\rfp\}\,\right)$ for the smooth configurations and
perturbations around them; $N=6$ and $r_g=1$.}
\label{tab:topgw}
\vspace{0.2cm}
\end{table}

\newpage

\appendix

\section{\bf Appendix}
\label{app}

In this Appendix we solve explicitly the minimization problem
~(\ref{oursapo}) when in the r.h.s. the Manton action is taken.
The result will be $\alpha^{\rm \pst min}(\alpha^\prime)$, and so
$U_\mu^{\rm\pst min}(U^\prime)=\exp(i\alpha_\mu^{\rm \pst
min}(\alpha^\prime))$.

Of course, due to gauge invariance, the fine configuration $\alpha_\mu(x)$
solving Eq.~(\ref{oursapo}) is not unique; indeed, an entire orbit of solutions
exists, all related by gauge transformations not altering the {\em fixed}
coarse configuration $\alpha_\mu^\prime(x_b)$.
It is possible to exploit this restricted freedom and fix the gauge
in order to simplify the problem.

We (partially) fix the gauge by requiring:

\be
\alpha^{\rm \pst min}_\mu(2x_b)\;=\;\alpha^{\rm \pst
min}_\mu(2x_b+\hat{\mu})\;\;,\;\;\;\;\;\;\;\;\;\mu=1,2\;\;.
\ee

\noi
In this gauge, the constraint:

\be
[\,\alpha_\mu(2x_b)+\alpha_\mu(2x_b+\hat{\mu})\,]\:({\rm mod}\;2\pi)\:=\:
\alpha^{\prime}_\mu(x_b)
\label{cond}
\ee

\noi
has the solution:

\be
\alpha^{\rm \pst min}_\mu(2x_b)\:=\:\alpha^{\rm \pst
min}_\mu(2x_b+\hat{\mu})\:=\:\f{1}{2}\,\alpha_\mu^{\prime}(x_b)
\:+\:n_\mu(x_b)\,\pi\;\;,
\label{sol1}
\ee

\noi
where $n_\mu(x_b)$ can be 0 or $\pm1$ (the sign depending on the sign of
$\alpha_\mu^{\prime}(x_b)$); exploiting the gauge freedom we can fix:
$n_\mu(x_b)=0$.

For a given lattice site of the blocked lattice $x_b$,
we define the block of the sites of the fine lattice belonging to $x_b$ as:
$\{x:\, x\,=\,2x_b+\lambda_1\hat{1}+\lambda_2\hat{2},\lambda_\mu=0,1,2\}$.
We see that in our particular gauge, the condition~(\ref{cond})
fixes the eight fine links lying on the border of the
above defined blocks (Eq.~(\ref{sol1})).
As a consequence, the residual minimizing problem decouples in the various
blocks; indeed the remaining degrees of freedom are the links {\em internal}
to the blocks (four for each block), and links internal to different blocks
do not communicate between themselves.

Minimization inside each block can be now trivially performed. Here gauge
invariance can be again exploited by fixing:

\be
\alpha^{\pst\rm min}_{\pst 1}(2x_b+\hat{2})\:=\:\alpha^{\pst\rm min}_{\pst
1}(2x_b+\hat{1}+\hat{2})\;\;.
\ee

\noi
The result is:

\bea
\alpha^{\rm \pst min}_{\pst 2}(2x_b+\hat{1})&=&\f{1}{4}(\ap_{\pst 2}(x_b)+
\ap_{\pst 2}(x_b+\hat{1}))\:+\:
(n_{\pst 1}(x_b)-n_{\pst 2}(x_b))\,\pi \;\;\;\;\nonumber\\
\alpha^{\rm \pst min}_{\pst 1}(2x_b+\hat{2})&=&\alpha^{\rm \pst min}_{\pst
1}(2x_b+\hat{1}+\hat{2})\:\:=\nonumber\\
\f{1}{4}(\ap_{\pst 1}(x_b)\!\!\!\!&+&\!\!\!\!\ap_{\pst 1}(x_b+\hat{2}))\:
+\:(n_{\pst 3}(x_b)+n_{\pst 4}(x_b)-n_{\pst 1}(x_b)-n_{\pst
2}(x_b))\,\f{\pi}{2}\;\;\;\; \nonumber\\
\alpha^{\rm \pst min}_{\pst 2}(2x_b+\hat{1}+\hat{2})&=&\f{1}{4}(\ap_{\pst
2}(x_b)+\ap_{\pst 2}(x_b+\hat{1}))\:+\:
(n_{\pst 4}(x_b)-n_{\pst 3}(x_b))\,\pi \;\;,\;
\label{sol2}
\eea

\noi
where the integer numbers $n_i(x_b), i=1,2,3,4$, are involved in the
consistency conditions

\be
F_{\pst 12}^{\rm \pst L}(2x_b)-n_{\pst 1}(x_b)\,2\pi\:\in\:(-\pi,\pi]\;\;,
\label{cons}
\ee

\noi
and analogous relations respectively for $F_{\pst 12}^{\rm\pst
L}(2x_b+\hat{1})$,
$F_{\pst 12}^{\rm\pst L}(2x_b+\hat{1}+\hat{2})$ and  $F_{\pst 12}^{\rm\pst
L}(2x_b+\hat{2})$,
where $F_{\pst 12}^{\rm\pst L}(x)$ is the lattice field tensor for the solving
fine configuration given by Eqs.~(\ref{sol1}) and~(\ref{sol2}).
For this solution, one has:

\bea
F_{\pst 12}^{\rm\pst L}(2x_b)\:-\:n_{\pst 1}\,2\pi\:=\:
F_{\pst 12}^{\rm\pst L}(2x_b+\hat{1})\:-\:n_{\pst 2}\,2\pi\:=\:F_{\pst
12}^{\rm\pst L}(2x_b+\hat{1}+\hat{2})\:-\:n_{\pst 3}\,2\pi\:=\nonumber\\
F_{\pst 12}^{\rm\pst L}(2x_b+\hat{2})\:-\:n_{\pst 4}\,2\pi\:=\:
\f{1}{4}F_{\pst 12}^{{\rm \pst L}\,\prime}(x_b)\:-\:
(n_{\pst 1}+n_{\pst 2}+n_{\pst 3}+n_{\pst
4})\,\f{\pi}{2}\;\;,\;\;\;\;\;\;\;\;\;\;
\label{rel}
\eea

\noi
where we indicate with $F_{\pst 12}^{{\rm \pst L}\,\prime}(x_b)$ the field
tensor of the coarse configuration $\alpha^{\prime}(x_b)$.
Using the above relations, its is clear that the total action of the
fine configuration, according to Manton's definition, reads:

\bea
\f{1}{2}\sum_x\,\left[\,F^{\rm\pst L}_{\pst 12}(x)\;({\rm
mod}\,2\pi)\,\right]^2\:=\;
\;\;\;\;\;\;\;\;\;\;\;\;\;\;\;\;\;\;\;\;\;\;\;\;\;\;\;\;\;\;\nonumber\\
\f{1}{8}\sum_{x_b}\,\left[\,F_{\pst 12}^{{\rm \pst L}\,\prime}(x_b)-
(n_{\pst 1}(x_b)+n_{\pst 2}(x_b)+n_{\pst 3}(x_b)+n_{\pst
4}(x_b))\,2\pi\,\right]^2\;\;.
\eea

\noi
The absolute minimum of the r.h.s. of the above equation among variations
of the $n_i(x_b)$'s is given by:

\be
S^{\min}\:=\:\f{1}{8}\sum_{x_b}\,\left[\,F_{\pst 12}^{\rm \pst
L\,\prime}(x_b)\,{(\rm mod}\,2\pi)\,\right]^2\;\;.
\label{minimum}
\ee

\noi
Observe that the set of integer numbers realizing the absolute
minimum~(\ref{minimum}) satisfy in particular the consistency
relation~(\ref{cons}) and the other three analogous relations
for $F_{\pst 12}^{\rm\pst L}(2x_b+\hat{1})$,
$F_{\pst 12}^{\rm\pst L}(2x_b+\hat{1}+\hat{2})$ and  $F_{\pst 12}^{\rm\pst
L}(2x_b+\hat{2})$. Eq.~(\ref{minimum}) is the Manton action for the coarse
configuration, apart from a factor $1/4$, which renormalizes the coupling
in a trivial way, $\beta\rightarrow\beta/4$.

For the fine minimizing configuration the following
relation holds (see Eq. (\ref{rel})):

\be
F^{\rm \pst L}_{\pst 12}(x)\;({\rm mod}\,2\pi)\:=\:\f{1}{4}\, \left[F^{\rm \pst
L \prime}_{\pst 12}(x_b)\;({\rm mod}\,2\pi)\right]\;,
\;\;\;x=2x_b+\lambda_{\pst 1}\hat{1}+\lambda_{\pst 2}\hat{2},\;
\lambda_{\mu}=0,1\;\;,
\ee

\noi
from which it is evident that the lattice topological charge
operator~(\ref{fpch}) reproduces itself under the RG transformation.

\end{document}